\documentclass[final,3p,times,twocolumn]{elsarticle}
\usepackage{color}
\usepackage[normalem]{ulem}
\usepackage{graphicx}
\usepackage{epsfig}
\usepackage{amssymb}
\usepackage{lineno}

\usepackage[hyperindex,breaklinks]{hyperref}
\usepackage{breakurl}

\biboptions{compress}

\newcounter{bla}

\newcommand{\Lag}{\mathcal{L}}

\journal{Computer Physics Communications}

\begin{document}

\begin{frontmatter}



\title{XQCAT: eXtra Quark Combined Analysis Tool}


\author[a,b,c]{D. Barducci}
\author[a,b]{A. Belyaev}
\author[d]{M. Buchkremer}
\author[c]{J. Marrouche}
\author[a,b]{S. Moretti}
\author[a,b]{L. Panizzi\corref{author}}

\cortext[author] {Corresponding author.\\\textit{E-mail address:} l.panizzi@soton.ac.uk}
\address[a]{School of Physics and Astronomy, University of Southampton, Highfield, Southampton SO17 1BJ, UK}
\address[b]{Particle Physics Department, Rutherford Appleton Laboratory, Chilton, Didcot, Oxon OX11 0QX, UK}
\address[c]{Physics Department, CERN, CH-1211, Geneva 23, Switzerland}
\address[d]{Centre for Cosmology, Particle Physics and Phenomenology (CP3), Universit\'e catholique de Louvain,\\ Chemin du Cyclotron, 2, B-1348, Louvain-la-Neuve, Belgium}

\begin{abstract}
% A submitted program is expected to be of benefit to other physicists or physical chemists, or be an exemplar of good programming practice, or illustrate new or novel programming techniques which are of importance to some branch of computational physics or physical chemistry.
% 
% Acceptable program descriptions can take different forms. The following Long Write-Up structure is a suggested structure but it is not obligatory. Actual structure will depend on the length of the program, the extent to which the algorithms or software have already been described in literature, and the detail provided in the user manual.
% 
% Your manuscript and figure sources should be submitted through the Elsevier Editorial System (EES) by using the online submission tool at \\
% http://www.ees.elsevier.com/cpc.
% 
% In addition to the manuscript you must supply: the program source code; job control scripts, where applicable; a README file giving the names and a brief description of all the files that make up the package and clear instructions on the installation and execution of the program; sample input and output data for at least one comprehensive test run; and, where appropriate, a user manual. These should be sent, via email as a compressed archive file, to the CPC Program Librarian at cpc@qub.ac.uk.

XQCAT (\textit{eXtra Quark Combined Analysis Tool}) is a tool aimed at determining exclusion confidence levels for scenarios of new physics characterised by the presence of one or multiple heavy extra quarks which interact through Yukawa couplings with any of the Standard Model quarks. The code uses a database of efficiencies for pre-simulated processes of QCD-induced pair production of extra quarks and their subsequent on-shell decays. 
In the version 1.2 of XQCAT the efficiencies have been computed for a set of seven publicly available search results by the CMS experiment. The input for the code is a text file in which masses, branching ratios and dominant chirality of the couplings of the new quarks are provided. The output of the code is the exclusion confidence levels of the test point for each implemented experimental analysis considered individually and, when possible, in statistical combination. 

\end{abstract}

\begin{keyword}
Extra Quarks \sep Vector Like Quarks \sep top partners \sep LHC  

\end{keyword}

\end{frontmatter}

% \linenumbers

% Computer program descriptions should contain the following
% PROGRAM SUMMARY.

{\bf PROGRAM SUMMARY}
  %Delete as appropriate.

\begin{small}
\noindent
{\em Manuscript Title:} XQCAT: eXtra Quark Combined Analysis Tool \\
{\em Authors:} D. Barducci, A. Belyaev, M. Buchkremer, J.~Marrouche, S. Moretti, L. Panizzi \\
{\em Program Title:} XQCAT \\
{\em Journal Reference:}                                      \\
  %Leave blank, supplied by Elsevier.
{\em Catalogue identifier:}                                   \\
  %Leave blank, supplied by Elsevier.
{\em Licensing provisions:}                                   \\
  %enter "none" if CPC non-profit use license is sufficient.
{\em Programming language:} Perl and C++ \\
{\em Computer:} Personal Computer \\
  %Computer(s) for which program has been designed.
{\em Operating system:} Any OS where Perl, C++ and ROOT are installed.\\
  %Operating system(s) for which program has been designed.
{\em RAM:} around 2.5 Mbytes                                              \\
  %RAM in bytes required to execute program with typical data.
{\em Number of processors used:} 1\\
  %If more than one processor.
{\em Supplementary material:}                                 \\
  % Fill in if necessary, otherwise leave out.
{\em Keywords:} Heavy extra quarks, vector-like quarks, top partners, LHC \\
  % Please give some freely chosen keywords that we can use in a
  % cumulative keyword index.
{\em Classification:} 11.1 General, High Energy Physics and Computing, 11.6 Phenomenological and Empirical Models and Theories \\
  %Classify using CPC Program Library Subject Index, see (
  % http://cpc.cs.qub.ac.uk/subjectIndex/SUBJECT_index.html)
  %e.g. 4.4 Feynman diagrams, 5 Computer Algebra.
{\em External routines/libraries:} C++, Perl and ROOT \\
  % Fill in if necessary, otherwise leave out.
{\em Subprograms used:} ROOT \\
  %Fill in if necessary, otherwise leave out.
% {\em Catalogue identifier of previous version:}*              \\
  %Only required for a New Version summary, otherwise leave out.
% {\em Journal reference of previous version:}*                  \\
  %Only required for a New Version summary, otherwise leave out.
% {\em Does the new version supersede the previous version?:}*   \\
  %Only required for a New Version summary, otherwise leave out.
{\em Nature of problem:} 
Experimental searches of new heavy extra quarks (XQs) usually consider minimal extensions of the Standard Model (SM) with only one XQ representation which couples only to the third generation SM quarks or only to the SM light quark families. In contrast, various theoretically-motivated scenarios of new physics predict a new quark sector, i.e. more than one new quark, with general couplings to SM quarks. Notable examples are composite Higgs models or universal extra-dimensions. Hence, recasting experimental limits for these scenarios can be a challenging task. To avoid time-consuming simulations and dedicated searches for scenarios which can already be excluded by current data, we present a tool for reinterpretation of existing experimental analyses, including those not made for XQs searches, in terms of exclusion confidence levels (eCLs) for XQ models. Our tool should serve as a useful preliminary approach to understand in a quick way the regions of validity of scenarios of new physics. \\
  %Describe the nature of the problem here.
%    \\
{\em Solution method:} The core of the tool consists in a database of pre-simulated efficiencies ($\epsilon$), defined as the ratio of signal events which survive a given set of experimental cuts over the total number of signal events. These efficiencies have been computed simulating the process of pair production and decay of XQs with masses in the range 400--2000 GeV and implementing the selection and kinematics cuts of a set of experimental searches at 7 and 8 TeV. The tool uses the database to reconstruct any scenario where XQs couple to SM quarks through Yukawa interactions and gives as output the eCL of the test point, characterised by values of the XQ masses and their branching ratios (BRs) into specific final states, for all the implemented searches.
It also provides eCLs for combinations of experimental searches when search bins are uncorrelated. The method has been validated against experimental analyses.
\\
  %Describe the method solution here.
   \\
% {\em Reasons for the new version:}*\\
  %Only required for a New Version summary, otherwise leave out.
%    \\
% {\em Summary of revisions:}*\\
  %Only required for a New Version summary, otherwise leave out.
%    \\
{\em Restrictions:} The efficiency database has been built under the following hypotheses: QCD-induced pair production of quarks with masses from 400 GeV to 2000 GeV with steps of 100 GeV; the electro-weak (EW) couplings of XQs with SM states have a dominant chirality, according to the hypothesis that new quarks are vector-like and that a new chiral generation is excluded with high confidence level (see main text for more details); in v1.2 the number of searches implemented is limited: four supersymmetric (SUSY) inspired CMS searches for final states with jets, missing transverse energy and variable number of leptons at 7 TeV [1-4], two SUSY-inspired CMS searches at 8 TeV (updates of two of the implemented 7 TeV searches) [5,6] and one CMS direct search at 8 TeV of a vector-like $t^\prime$ (a top quark partner) coupling to third generation SM quarks [7]. The code also relies on the possibility of reproducing experimental results by using a cut-and-count analysis technique and applying a eCL procedure.\\
  %Describe any restrictions on the complexity of the problem here.
   \\
% {\em Unusual features:}\\
  %Describe any unusual features of the program/problem here.
%    \\
% {\em Additional comments:}\\
  %Provide any additional comments here.
%    \\
{\em Running time:} Around 30 seconds with two XQs and with 100k toy Monte Carlo (MC) experiments yelding specific numbers of signal and background events to determine the eCL (tested on a DELL Precision M4700 laptop with a Kubuntu Linux 12.04 distribution). \\
  %Give an indication of the typical running time here.
   \\

% * Items marked with an asterisk are only required for new versions
% of programs previously published in the CPC Program Library.\\
\end{small}

\section{Introduction}
\label{sec:introduction}
Different classes of new physics models predict the existence of new heavy extra quarks (XQs). While the existence of a fourth chiral generation has been excluded with high confidence level under the assumption of a Standard Model (SM)-like Higgs boson, vector-like quarks cannot be excluded due to their decoupling property. Such new top partners can naturally exist near the EW scale without upsetting the existing measurements, and are often crucial to cure the divergences in the Higgs mass corrections in scenarios going beyond the Standard Model (BSM). Notable examples involve models with composite Higgs states~\cite{Dobrescu:1997nm,Chivukula:1998wd,He:2001fz,Hill:2002ap,Agashe:2004rs,Contino:2006qr,Barbieri:2007bh,Anastasiou:2009rv,Gripaios:2014pqa}. They also appear as manifestations of extended symmetries in models of Little Higgs~\cite{ArkaniHamed:2002qx,Yang:2014mba} and in Grand Unified scenarios~\cite{Rosner:1985hx,Robinett:1985dz} or as excitation modes of the SM quarks in theories with extra 
dimensions~\cite{Antoniadis:1990ew,Appelquist:2000nn,Csaki:2003sh,Cacciapaglia:2009pa}. New heavy quarks have also been the subject of intense experimental efforts, that resulted in a large number of analyses aimed at putting bounds on their masses under specific hypotheses on their nature~\cite{twikiATLAS,twikiCMS}. Experimental searches on new heavy XQs, either chiral or vector-like, are usually done under specific assumptions about the couplings of the new quarks to SM states. To date, the general assumption is that new quarks predominantly couple to either the first or the third generation of SM quarks and therefore kinematics cuts are tuned to be sensitive to specific final states. In addition, bounds on XQs masses are set assuming the existence of only one new quark besides the SM states and obtained for different values of their branching ratios (BRs). 
Considering a simple extension of the SM with a single vector-like quark representation besides the SM states, the allowed physical states are limited, so that vector-like quarks can only appear with four different charges: two standard partners of SM quarks $t^\prime$ and $b^\prime$ (with  electric charges $+2/3$ and $-1/3$, respectively) and two exotic quarks $X$ with charge $+5/3$ and $Y$ with charge $-4/3$, respectively (see, e.g.,~\cite{delAguila:2000rc,Cacciapaglia:2012dd,Okada:2012gy}). It is possible to build scenarios with more exotic vector-like quarks, but they will not interact directly with SM quarks due to the large charge gap. Thus, they will undergo chain-decays (i.e. decay into each other) or a three-body decay, to reach a final state composed of SM particles. For the most recent experimental bounds for vector-like $t^\prime$, $b^\prime$, $X$ and $Y$ quarks we refer to the ATLAS and CMS public results Web pages~\cite{twikiATLAS,twikiCMS}.

Constraining BSM scenarios that predict the existence of a new sector of quarks requires a reinterpretation of experimental data, which often is not limited to a rescaling of experimental bounds to account for different masses or BRs. The presence of more quarks that contribute to the same final state and the possibility of having interactions not accounted for by the experimental analysis make the determination of allowed and excluded regions in the parameter space of the scenario under consideration quite challenging. Conversely, performing full simulations to determine these regions limits the possibility to effectively scan on the parameter space of the model considered. The purpose of XQCAT is to provide the high energy physics community with a fast and reliable method to determine in a conservative way (see Sec.~\ref{sec:restrictions}) the excluded parameter regions of a generic model
with several XQs of different type. 

The structure of the framework can be summarised in the following points (a more detailed description is provided in the following sections).
\begin{enumerate}
\item Monte Carlo (MC) simulations have been performed for processes of QCD-induced pair production and on-shell decays of XQs with different masses and considering all possible decay channels allowed by the assumption of Yukawa mixing with SM quarks. 
\item The selection and kinematics cuts of a number of experimental searches have been implemented and applied to the signal obtained with the MC simulation to create efficiency database entries for each quark, each mass and each possible channel. The efficiencies $\epsilon$ are defined as the ratio of signal events which pass the cuts with respect to the total number of signal events (minimum-bias hypothesis), and they also include detector acceptance effects.
\item After an appropriate validation, the efficiency database, together with a cross section repository, has been included in the public tool, which then has only to apply simple analytical relations to obtain the number of signal events that survive the experimental cuts for the provided input, which only consists of masses and BRs of the XQs present in the benchmark point under consideration. The number of signal events is given by the simple relation $S=\sigma~\Lag~\mathrm{BR}_{Q\to q \mathcal{B}}~\mathrm{BR}_{\bar Q\to \bar q \mathcal{B}}~\epsilon_{Q\bar Q\to q \bar q \mathcal{B}\mathcal{B}}$, where $\sigma$ is the cross section for pair production of the XQ $Q$, $\Lag$ is the integrated luminosity, $q$ are SM quarks and $\mathcal{B}$ is a SM boson ($W$, $Z$ or $H$).
\item Finally, considering the public experimental data about observed events, background, and uncertainties, the tool computes the exclusion confidence level (eCL) of the benchmark analysed. The eCL is obtained by using a Monte Carlo procedure to run a series of toy experiments. For each of these toy experiments, the mean value of the Poisson distributions of signal and background events follows a Gaussian distribution with uncertainties provided by the user (for the signal) and by experimental data (for the background). The eCL is then computed through a $CL(s)$ procedure~\cite{Read:2000ru,Read:2002hq} by assuming that the signal and background distributions are Poissonian and centered on the values determined by the toy experiments.
\end{enumerate}

The timescale for obtaining the output of XQCAT depends on the number of heavy quarks in the input benchmark, on the number of combinations for their decays and on the level of accuracy the user would like to achieve. Higher accuracies are achieved by increasing the number of iterations of the MC analysis which determines the eCL. A run can therefore take from few seconds to some minutes increasing with the number of input states, decay channels and/or iterations. The public part of the code that computes the number of signal events for the input provided has been written in Perl and does not require any further external library. The limit code, which computes the eCL through a simple MC analysis, has been written in C++ and requires a basic ROOT~\cite{Root} installation (the code has been tested on a ROOT 5.34.09 version). The XQCAT code therefore requires basic Perl, C++ and ROOT, and does not need to be installed: it can be used out-of-the-box by setting the parameters in the input and initialisation 
cards and 
by running the main executable. The non-public part of the project consists in the MC simulations of the pair production and decay processes and on the 
subsequent hadronisation, detector simulations and 
efficiency extraction. Storing all the kinematical information of the simulated events for all channels and all masses is not 
practically feasible, while storing the efficiencies is by far more convenient in terms of memory size, as the database is composed of text files.

There are other recasting tools on the market, as SModelS~\cite{Kraml:2013mwa,Kraml:2014sna} and Fastlim~\cite{Papucci:2014rja}. These tools are similar in concept to XQCAT, but they are dedicated to testing SUSY scenarios, and therefore they are complementary to XQCAT in terms of the new physics scenarios they are able to constrain.

Other publicly available tools are in principle able to perform the same analysis as XQCAT, but with different characteristics. CheckMATE~\cite{Drees:2013wra,Kim:2015wza} and MadAnalysis~\cite{Conte:2012fm,Dumont:2014tja,Conte:2014zja} accept simulated events and apply the selection and kinematic cuts of a large number of analyses already present in their database. The main difference between XQCAT and these tools is the possibility (in XQCAT) to quickly perform scans over the parameter space of complex scenarios characterised by the presence of a large number of XQs: XQCAT already contains a database of efficiencies for pre-simulated events, and it simply sums the number of signal events for each XQ present in the model in each channel, and reconstruct the total signal by algebraic procedures. This means that the user does not need to run event simulations for each model benchmark they wish to test, as is the case for CheckMATE and MadAnalysis5; full simulations can be performed to test only the most 
interesting regions which can be determined in a conservative way through XQCAT. One further difference is the set of experimental searches contained in the database of CheckMATE and MadAnalysis. Although the number of search analyses implemented in these tools is currently greater than in XQCAT, the emphasis is different: almost all the searches implemented in those tools are SUSY-inspired, and therefore require large amounts of missing transverse energy (MET) in the final state. The presence, in XQCAT, of a search dedicated to the detection of vector-like top partners makes the current version of XQCAT more suitable to test the scenarios we are interested in. Nevertheless, in future developments of XQCAT we are planning to interface our simulation chain with their efficiency extraction codes, in order to exploit larger databases and the possibility to implement new searches (more specific for our purposes) in their frameworks.

This paper serves as a description of the structure of the full project, including the non-public modules used to determine the efficiency database, and it is the instruction manual of XQCAT. The code has 
already been used to produce physical results in a previous analysis~\cite{Barducci:2014ila}, where further details on validation steps and on the structure of the code can also be found. The modules that extract the efficiencies have also been used previously to produce physical results, in Ref.~\cite{Buchmueller:2013exa}, where further details on the validation of this part of the project can be found.

The code is publicly available for download and use at the following address (Ref.~\cite{XQCATwebpage}):\\
\centerline{\tt \href{https://launchpad.net/xqcat}{https://launchpad.net/xqcat}}.

\section{Signal generation}
\label{sec:simulationdetails}

XQCAT contains a database of efficiencies obtained by applying the same cuts considered in experimental analysis to simulated signal events. In this section, a detailed description of the simulation steps and of the underlying assumptions is provided.

\subsection{Assumptions on the XQs}
The basic assumptions about the properties of the XQs are as follows.
\begin{itemize}
\item Their QCD interactions are exactly the same as for SM quarks.
\item They couple to SM quarks and with the SM Higgs boson through Yukawa interactions. Therefore, the only allowed representations for the XQs are singlets, doublets or triplets~\cite{delAguila:2000rc,Cacciapaglia:2012dd,Okada:2012gy}. Couplings to all generations of SM quarks are allowed. We do not consider couplings with other exotic vector or scalar states. The general Lagrangian term which describes the interaction between an XQ representation $Q$ and any of the SM quarks $q_i$ is
\begin{itemize}
\item \textit{extra quark singlet}
\begin{equation}
 \Lag_{int}=-\lambda_{L,R}^{Qi} \bar{q}^i_{R,L} H^{(c)} Q_{L,R} + h.c. 
\label{SingletCoupling} 
\end{equation}
\item \textit{extra quark doublet}
\begin{equation}
 \Lag_{int}=-\lambda_{L,R}^{Qi} \bar{Q}_{R,L} H^{(c)} q^i_{L,R} + h.c. 
\label{DoubletCoupling} 
\end{equation}
\item \textit{extra quark triplet}
\begin{equation}
 \Lag_{int}=-\lambda_{L,R}^{Qi} \bar{q}^i_{R,L} \tau^a H^{(c)} Q^a_{L,R} + h.c. 
\label{TripletCoupling} 
\end{equation}
\end{itemize}
where $i=1,2,3$, $\tau^a$ are the SU(2) Pauli matrices and the $\lambda^{Qi}$'s are the corresponding Yukawa couplings. The latter quantities are assumed to be real parameters: while new non-zero complex phases may be present for XQ mixing with the SM quarks, they are expected to play a minor role in the present phenomenological analysis and will be neglected in the following.
The requirement of standard Yukawa couplings with the SM quarks also limits the charge of the XQs considered in the simulation: they can have charges $-4/3$, $-1/3$, $+2/3$, $+5/3$. If no other new state is present in the model, quarks with more exotic charges cannot interact directly with SM quarks and can only decay to SM states through three-body or chain decays to other XQs (e.g., $Q_{8/3}\to Q_{5/3}W^+ \to t W^+W^+$). Other possible couplings have not been considered in the first version of the code: for example XQs may couple to SM quarks and to new neutral scalars or vectors via
\begin{eqnarray}
  \Lag_{int}^S &=& -\lambda_{L,R}^{QiS} \bar{Q}_{R,L} S^0 q^i_{L,R} + h.c. \\
  \Lag_{int}^V &=& -\lambda_{L,R}^{QiV} \bar{Q}_{L,R} \gamma^\mu V^0_\mu q^i_{L,R} + h.c. \,.
\end{eqnarray}
These interactions may be possible in models where the XQs couple to a dark matter candidate (as in Universal Extra Dimensions) and they will be included in future upgrades of the code.
\item Assuming a SM Higgs field, Eqs.~\ref{SingletCoupling},~\ref{DoubletCoupling} and~\ref{TripletCoupling} imply that the Yukawa and the gauge couplings of the XQs to $W$, $Z$ and $H$ are (predominantly) chiral. According to this statement, proven in Refs.~\cite{delAguila:2000rc,delAguila:1982fs,AguilarSaavedra:2009es,DeSimone:2012fs,Buchkremer:2013bha} for scenarios with one and multiple quark representations, the chiralities of the XQ will be fixed at the production level to be either left- or right-handed.
\item The EW couplings of the XQs are assumed to be small enough that it is possible to factorize QCD production from on-shell decays into SM quarks and $W$, $Z$ or $H$ bosons in the narrow width approximation. This assumption is generally well satisfied in most theoretical models predicting a new quark sector.
\end{itemize}

\subsection{Simulation details}
The simulations have been performed for QCD-induced pair production and on-shell decays of the XQs, also including QCD initial- and final-state radiation appearing as separate jets $j$,
\begin{equation}
 p p \to Q \bar{Q} + \{0,1,2\}\,j
 \label{eq:pairproduction}
\end{equation}
The possible decays of the XQs depend on their charge and the complete list is as follows (with the charge of the XQs in decreasing order):
\begin{eqnarray}
\begin{array}{lcl}
 X &\to& W^+ u_i \\
 t^\prime &\to& W^+ d_i, Z u_i, H u_i \\
 b^\prime &\to& W^- u_i, Z d_i, H d_i \\
 Y &\to& W^- d_i
\end{array}
\end{eqnarray}
with $i=1,2,3$. The above decays amounts to 24 distinct partonic states when summed over the 3 families. However, jets generated by light quarks (including charm) are hardly distinguishable and hence $u_i=\{j, t\}$ and $d_i=\{j, b\}$, giving a reduced total of 16 distinguishable final states.

The tools used for the MC simulation are:
\begin{itemize}
\item \textit{MadGraph5 v.1.5.8}~\cite{Alwall:2014hca}: dedicated models in the MadGraph v4 format have been implemented for each particle, considering couplings of each chirality and the models have been validated against the UFO~\cite{Degrande:2011ua} models in Ref.~\cite{feynrulesVLQ} implemented through the FeynRules package~\cite{Christensen:2008py,Alloul:2013bka}. This tool has been used to simulate the QCD-induced pair production process of Eq.~\ref{eq:pairproduction}. The kinematic cuts at partonic level are those in the standard MadGraph \verb#run_card#; this is justified by the fact that the default cuts are looser than the selection and kinematics cuts of the experimental analysis implemented in the tool, and therefore they do not remove signal events which would be relevant for the determination of the experimental efficiencies. Simulation of QCD radiation has been included, but since the jet merging procedure involves the interface between MadGraph and Pythia, the procedure and related 
parameters are described 
in 
the dedicated Sect.~\ref{sec:jetmatching}. The simulated events have been 80k for each channel (split in two runs of 40k events each) to allow for enough statistics (see Sec.~\ref{sec:effdatabase} for more details about checks on the statistics of the simulation). The parton densities used for the simulation are the CTEQ6L1~\cite{Pumplin:2002vw} and the scale factors (renormalisation and factorisation scales) are set automatically by MadGraph on an event-by-event basis: the default MG5 algorithm sets both scales on the central $m_T^2$ scale of the event, which for the case under consideration (pair production of heavy particles) corresponds to the geometrical mean of $M^2+p_T^2$ for each particle~\cite{MGscales}. The reason for using a MadGraph v4 model and validating it against the UFO~\cite{Degrande:2011ua} implementations~\cite{feynrulesVLQ} is practical and related to the tool we use for the simulation of the decay chains (see BRIDGE below). 

\item \textit{BRIDGE v.2.24}~\cite{Meade:2007js}: The decays of the XQs are simulated through BRIDGE, which adopts a MC procedure for the decays. XQs are assumed to be produced on-shell and then decayed. When separating production from decay, however, information about the spin correlation between quark and antiquark is lost. To preserve it, one should perform the matrix-level simulation of the full process down to the final state instead of separating production from decay, but this is not practically feasible because too computationally intensive. BRIDGE applies an approximation to account for spin correlations without explicit spin information, and this is been verified as satisfactory in Ref.~\cite{Meade:2007js}. BRIDGE is also used to decay the heavy SM states to the lightest ones. The decays of the XQs $Q$ and $\bar Q$ are performed separately, considering the quark and the antiquark as independent particles. By assigning 100\% BRs to different channels for quark and antiquark independently, it is 
possible to perform the simulation independently for each channel in order to obtain the corresponding efficiencies. Of course this results in the simulation of unphysical scenarios, like $t^\prime \bar t^\prime \to t Z \bar t H$ with 100\% BR of $t^\prime$ into $t Z$ and 100\% BR of $\bar t^\prime$ into $\bar t H$. However, this will allow the rescaling of the cross section corresponding to BRs in the final code.

\item \textit{Pythia v.6.426}~\cite{Sjostrand:2006za}: this is used to simulate initial- and final-state radiation (via the parton shower formalism) and hadronisation of the partonic final state. The used flags are those of the default \verb#pythia_card# in MadGraph. In detail: parton shower flags MSTP(61) for QCD initial state radiation and MSTP(71) for QCD and QED final state radiation have been set to 1 and the MSTJ(1) flag corresponding to the fragmentation code has been set to 1, i.e. string fragmentation according to the Lund hadronisation model (more details about these flags can be found in the Pythia manual~\cite{Sjostrand:2006za}). As a check of the consistency of the switches, we have verified that the cross section after Pythia (which can be retrieved at the end of the Pythia log file) corresponds within 20\% to the cross section of the $2\to2$ process of the matrix-element generator (MadGraph). To perform the jet merging procedure, specific flags have been set, and we refer to Sect.~\ref{sec:
jetmatching} for more details.

\item \textit{Delphes v.2.0.2}~\cite{Ovyn:2009tx}: detector simulation has been performed using dedicated cards for ATLAS and CMS, with suitable modifications from the default Delphes cards to account for more accurate $b$-tagging. The reason for using Delphes2 instead of Delphes3~\cite{deFavereau:2013fsa} is purely historical: the framework has been used and validated in Ref.~\cite{Buchmueller:2013exa} and the code itself has been customised and fully validated against experimental data. However, future updates of the efficiency database will rely on an upgraded framework with a Delphes3 implementation.

\item \textit{Efficiency code, not public} (see Ref.~\cite{Buchmueller:2013exa} for further details): the outcome of the simulation, i.e. ROOT files from Delphes with full kinematic information about the signal events, is then passed to the efficiency-extraction code, in which the selection and kinematic cuts of the implemented experimental searches have been reproduced~\cite{Buchmueller:2013exa}. This code is written in C++ and consists essentially in functions for the identification of final states and selection of events which survive the implemented cuts. The list and properties of the implemented searches, together with details about their validation, are provided in Sects.~\ref{sec:implementedsearches} and \ref{sec:validation}.
\end{itemize}

\begin{figure*}[h!]
\begin{center}
\bf XQCAT project structure\\
\epsfig{file=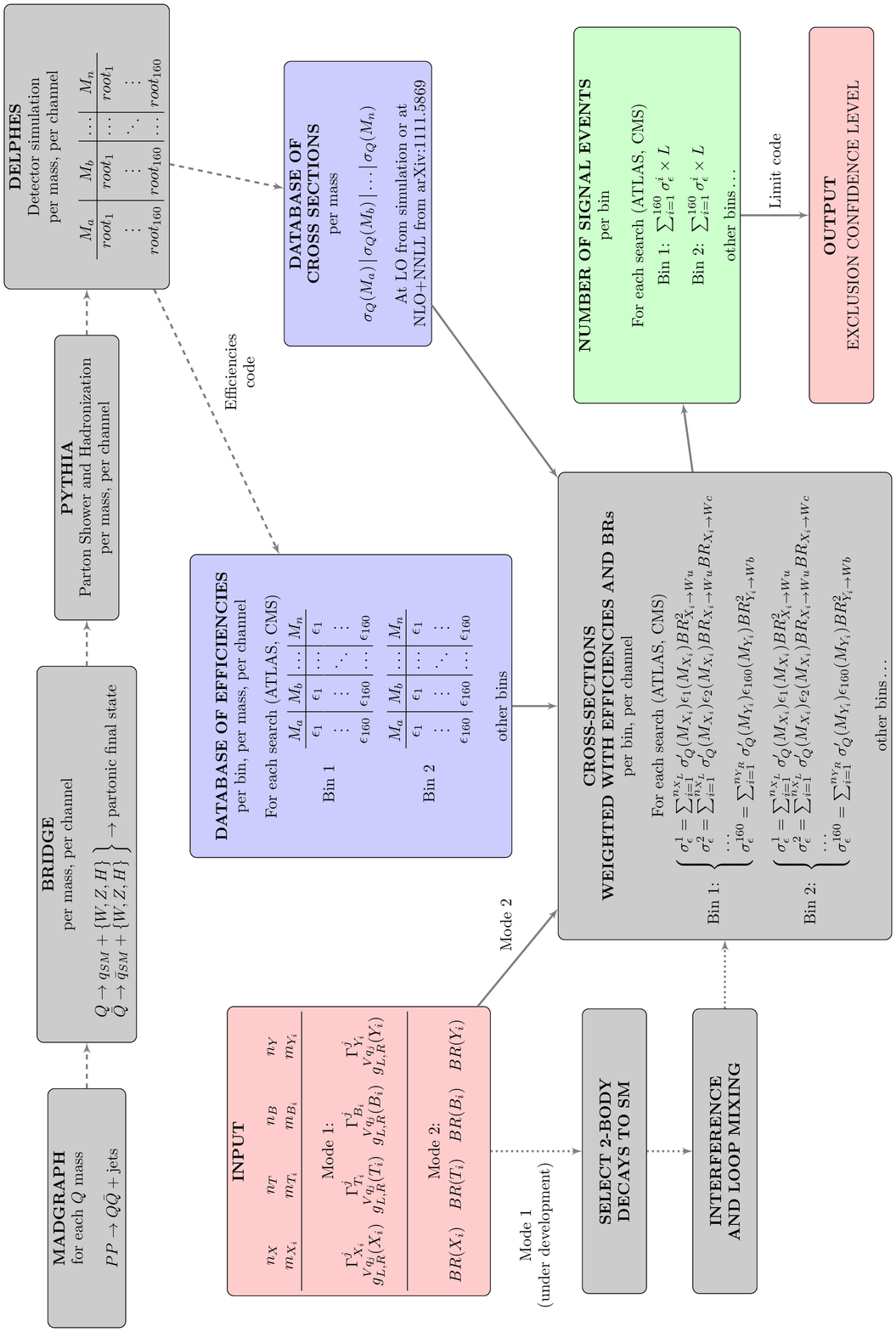,height=0.89\textheight,angle=0}
\end{center}
\caption{Flowchart of the XQCAT project. Modules connected by solid lines are part of the public code. Dashed connections correspond to the non-public sections. Dotted connections correspond to modules which are currently under development and will be published in a code update. Grey modules correspond to internal or off-line computation, pink modules correspond to the input and output of the code, light-blue modules correspond to databases, while the green module correspond to a partial output: the number of signal events can be retrieved and used for custom analyses.}
\label{fig:flowchart}
\end{figure*}

The full flow of the simulation can be found in Fig.~\ref{fig:flowchart}. The total number of simulations (for each mass) corresponds to the full set of combinations of decay channels for the XQs. Since decays of particles and antiparticles are treated independently, the total number of simulated channels are $2\times2+2\times2+6\times6+6\times6=80$, and this number has still to be multiplied by 4 because simulations have been performed for both coupling chiralities and for two different LHC energies (7 TeV and 8 TeV), for a total of 320 simulations for each quark mass.

\subsection{Jet merging procedure}
\label{sec:jetmatching}

Additional jets from initial- and final-state QCD radiation can play a relevant role in the determination of the kinematics of the scattering process. If the shapes of kinematic distributions are modified by the presence of further jets, the efficiency of kinematic cuts change and therefore the number of signal events may increase or decrease. The inclusion of QCD radiation in the simulation process can be done both at the matrix-element level (MadGraph) and at the parton-shower level (Pythia). The former is more suitable for modeling hard and well separated jets, the latter is more effective in the description of soft and collinear radiation. A consistent treatment of QCD radiation can be done by adding jets both at the matrix-element level and at the parton-shower level, and by using an algorithm to merge the two methods and remove double counting in the overlapping region. 
Within the MadGraph-Pythia flow it is possible to apply different jet-merging algorithms by setting the values of some dedicated parameters and flags. The description of the jet-merging procedure can be found in Refs.~\cite{Alwall:2007fs,Alwall:2008qv} and in the MadGraph wiki pages~\cite{MGmatch1,MGmatch2}. In the following we will describe our settings for the processes we have simulated in the development of XQCAT.

At the MadGraph level, the jet-merging parameters to set are in the run\_card: 
\begin{itemize}
\item the switch \verb#ickkw#, which selects the merging algorithm, has been set to 1, corresponding to MLM merging~\cite{Mangano:2006rw};
\item the switch \verb#ktscheme#, which selects the algorithm to cluster final state partons into jets, has been set to 1, corresponding to the Durham $k_\bot$ algorithm~\cite{Brown:1991hx,Catani:1993hr};
\item the \verb#xqcut# parameter (minimal distance in phase space between partons for the matrix-element calculation) has been set to different values depending on the XQ mass. The ``best'' value of \verb#xqcut# has been chosen by checking the smoothness of the differential jet rate distributions, as indicated in Ref.~\cite{Alwall:2008qv}, for each XQ mass we probed.
\end{itemize}
All other jet-merging parameters have been kept to the default values in the run\_card.

At the Pythia level, the jet-merging parameters can be found in the pythia\_card:
\begin{itemize}
\item the Pythia MSTP(81) flag, which sets how showers are ordered has been set to 20, corresponding to $p_\bot$-ordered showers;
\item the MadGraph5 SHOWERKT flag, which applies the shower-$k_\bot$ scheme described in~\cite{Alwall:2008qv} has been set to true;
\item the MadGraph5 QCUT parameter has been set to different values depending on the mass of the XQ. However, by using the shower-$k_\bot$ scheme, the QCUT value can be set equal to the \verb#xqcut# parameter (see Ref.~\cite{Alwall:2008qv}). A table of the \verb#xqcut#(=QCUT) values for each XQ mass can be found in the XQCAT webpage~\cite{XQCATwebpage}.
\end{itemize}
All other Pythia switches have been kept to their default values.

\section{Construction of the efficiency database}
\label{sec:effdatabase}

The efficiencies are defined as the ratio of signal events which survive a given set of experimental cuts over the total number of signal events. They are computed for each experimental bin and they include acceptance effects of the detector simulation.
Efficiencies computed with the simulation are stored in a folder contained in the XQCAT core (see below,~\ref{app:structure}). Efficiencies are sorted according to the XQ mass and its 
charge and also according to the experiment (ATLAS and CMS, even if in the first version of XQCAT only CMS analyses have been implemented and validated) and the LHC energy at which the 
experimental analyses have been done (7 or 8 TeV, in the first version of XQCAT). A text file containing the experimental efficiencies corresponds to each simulated channel, for each 
experiment and for each LHC energy. Every file contains the efficiencies for each bin of all the implemented analyses. Efficiencies are contained within tags which identify the 
experimental analysis, and are stored as relative numbers (not percentages, i.e. 1\% is stored in the database as 0.01). A typical efficiencies set contained in one of such files 
appears as\footnote{The example corresponds to the channel $ZtZ\bar{t}$ for a predominantly left-handed $t^\prime$ at 400 GeV and for the CMS analysis B2G-12-015 at 8 TeV~\cite{Chatrchyan:2013uxa}. Note that the newline before the last numerical value has been inserted just for a matter of presentation.}:\\\\
\begin{footnotesize}
$<$CMS8\_B2G12015$>$\\ 
5.12482917982918e-05~0.00201130951823688\\0.00313643032084553\\
$<$/CMS8\_B2G12015$>$\\\\
\end{footnotesize}
where the numbers correspond to the efficiencies for the 3 implemented bins of this specific search (see Sec.~\ref{sec:implementedsearches} for more details.). 

For all experimental searches
certain bins have a very low signal efficiency, while others are more important for the determination of the eCL. 
The importance of the bins is determined by computing the significance $S/\sqrt{S+B}$ for  each bin 
with $S=\sigma L\epsilon$ (where $\epsilon$ is the efficiency, and $L$ is the integrated luminosity) and the background $B$ given by the experiment for that bin. 
We define bins as relevant if  their  significance is larger than the median of the distribution of the significances of all bins in that search. 
In order to determine the number of the cross section-weighted signal events
with an accuracy  better than 20\%, a requirement is placed on the number of unweighted events\footnote{In this context, unweighted events are all events that fall on the relevant bins before weighting them with cross section and luminosity.} for each relevant bin.
The  20\% precision is achieved by requiring 25 unweighted event in each relevant bin:
assuming events follow a Poissonian distribution, the error on the number of events in a bin scales as $N^{-1/2}$, 
so with 25 events or more the relative uncertainty on the efficiency is below 20\%.
We have performed additional simulations to guarantee that at least 
half of the relevant bins contain more than 25 events. 
As a further remark, it is useful to notice that in the XQCAT framework all the background yields and uncertainties are those reported by the experiment and, as such, each bin in the eCL calculation are treated as independent. This means that we do not consider a correlated background estimate. If we did, ignoring a bin where the signal is zero would not be correct, because that bin still contains a measurement of background and data, and can be used as an estimate of the background to further constrain across all bins.

\section{Determination of the eCL}
\label{sec:eCLdetermination}

The analytical and computational issues related to the determination of the eCL and on how to interpolate values between simulated points have been extensively discussed in Appendix A of Ref.~\cite{Barducci:2014ila} and we refer the reader to this reference for more details. In this section, the purely code-related part of the eCL computation is discussed. The limit code  (see~\ref{subsec:limitcode}) performs a loop over the number of user-defined iterations and over the bins of the considered experimental analysis (or combination of analyses, if not correlated) and builds Poissonian distributions for background and signal+background. Specifically, for each bin and in each iteration, the mean values $\lambda_s$ and $\lambda_b$ of the Poissonian distributions are themselves Gaussian-distributed around the central values of signal and background events, respectively, with standard deviations given by the uncertainties on the signal and background events themselves. This procedure allows for the construction of 
the likelihood ratio test-statistic $Q$ for a given experimental result $\vec X$~\cite{Read:2000ru,Read:2002hq}:
\begin{equation}
 Q = \frac{\Lag(\vec X, s+b)}{\Lag(\vec X,b)} = e^{-\sum_{i=1}^{N_{\mathrm{bins}}}s_i} \prod_{i=1}^{N_{\mathrm{bins}}} \left(1+\frac{\lambda_{s,i}}{\lambda_{b,i}}\right)^{n_i},
\end{equation}
where $n_i$ is the number of observed events.
To define  the  eCL for a given scenario with vector-like quarks, we use p-values\footnote{The p-value is the probability of finding a value of the test statistics as or more extreme than the observed value, under the assumption of a true null hypothesis.} for the two hypotheses
-- with and without signal -- $p(s+b)=1-CL(s+b)$ and $p(b)=1-CL(b)$, respectively, where $CL(s+b)$ and $CL(b)$ are corresponding confidence levels. Then eCL is given by the following relation:
\begin{equation}
 \mathrm{eCL} \equiv 1-CL_s = 1-\frac{CL(s+b)}{CL(b)} = 1 - \frac{1 - p(s+b)}{1-p(b)} 
\end{equation}
which can be found  by  integrating the distributions of the test-statistics determined above.
The 1, 2 and 3 $\sigma$ exclusions correspond to value of the eCL of approximately 0.68, 0.95 and 0.9973, respectively.

It can be useful to notice that the limit code can be used as an independent module to determine the eCL for any set of signals, backgrounds (with relative uncertainties) and observations for any number of uncorrelated search bins. To allow the user to access the limit code as a separate piece, in the same folder, a 
self-contained limit code file is provided, which can be customised and run by compiling it with the provided \verb#makefile#.

\section{Implemented searches}
\label{sec:implementedsearches}

All data exploited by XQCAT are presently public and we intend it to remain so also in the future, irrespectively of the experimental source (ATLAS or CMS). In the current version of the code, seven searches have been implemented: four at 7 TeV and three at 8 TeV, all by CMS. However, for each specific search, not every bin has been considered for the extraction of the efficiencies or for the computation of the eCLs.
All details about searches and their implementation are in the following list (that can be also found in Ref.~\cite{Barducci:2014ila}).

\begin{itemize}
\item \underbar{\textit{Direct search of XQs}} We accounted in XQCAT for the CMS analysis associated to the B2G-12-015 note~\cite{Chatrchyan:2013uxa}, at $\sqrt{s}=8$ TeV with a 19.5 fb$^{-1}$ luminosity sample, for a pair produced $t^\prime$ quark that mixes only with SM quarks from the third generation and can decay to $W^+b, Zt$ or $Ht$ with variable (user-defined) BRs. In the aforementioned publication, CMS extract the 95\% eCL lower limits on the $t^\prime$ quark mass for different combinations of its BRs using six alternative channels: two single lepton (single electron and single muon), three di-lepton (2 opposite-sign and 1 same-sign) and one tri-lepton channel, all containing tagged $b$-jets in the final state. No deviations from the SM predictions  were observed when considering a large number of benchmark points with different  BRs. In a cut-and-count approach the sensitivity of the search is largely due  to the multi-lepton channels, while the single lepton channels require a more 
sophisticate treatment for the analysis (the so-called BDT discriminants). The limits for the multi-lepton channels only can be found in the wiki page of the search~\cite{twikiB2G12015} and the quoted observed bounds are in the range $592\textendash794$ GeV, depending on the assumed BRs. Among the channels available in the search we have implemented one opposite-sign, one same-sign and one tri-lepton channel, see Sec.~\ref{sec:effvalidation} for more details.
\item \underbar{\textit{SUSY searches}} XQCAT implements four searches inspired by scenarios potentially induced by
SUSY, each characterised by the presence of a different number of electrons/muons in the final state and large missing transverse energy: 0-lepton ($\alpha_{T}$)~\cite{Chatrchyan:2012wa}, single lepton ($L_{p}$)~\cite{Chatrchyan:2012sca}, opposite-sign dilepton (OS)~\cite{Chatrchyan:2012te} plus same-sign dilepton (SS)~\cite{Chatrchyan:2012sa}, considering the entire 4.98 fb$^{-1}$ 2011 dataset at $\sqrt{s}=7$ TeV. We also have accounted for the updated $\alpha _{T}$ analysis~\cite{Chatrchyan:2013lya} and same-sign~\cite{Chatrchyan:2012paa} search at 8 TeV, based on 11.7 fb$^{-1}$ and 10.5 fb$^{-1}$ of luminosity, respectively. It has also been checked that the selected searches are uncorrelated and therefore it is possible to statistically combine them without the need of a correlation matrix, thereby yielding 95\% eCL bounds at 7 TeV (combination of 4 searches), 8 TeV (combination of 2 searches) and 7+8 TeV (combination of 6 searches). The validation of the implementation of these searches can be found in 
Ref.~\cite{Buchmueller:2013exa}, to which we refer the reader for details.
\end{itemize}

\section{Code validation}
\label{sec:validation}

Analogously to the eCL determination section, detailed discussions about the code restrictions and the validation procedure have been provided already in Ref.~\cite{Barducci:2014ila}, to which we refer for further details, while in this section we will just summarise them.

\subsection{Code limitations}
\label{sec:restrictions}

It order to establish a robust and conservative exclusion limit it is important to  consider all the effects, not included in our calculation, that can affect the computation of the eCLs, and in particular it is important to identify those that can reduce the number of predicted signal events.
However, also an over-conservative estimate would not be appropriate as it would give rise to too weak a bound, making therefore important to consider also the effects that can increase the final event rate.
The main factors that could in principle affect the calculation of the eCLs are the following.
\begin{itemize}
\item \emph{Mass point interpolation}: having computed the efficiencies just for a limited number of XQ masses, at step of 100 GeV, when computing eCLs for masses between simulated values we need to adopt some interpolation procedures. We refer to Appendix A of Ref.~\cite{Barducci:2014ila} for a detailed description and comparison of the various methods.
\item \emph{Chain decays between XQs}: we have not included chain decays like $Q\to Q^\prime V, V=W,Z,H$ in our calculation in order to keep the tools simple. Moreover, even when these chain decays are kinematically allowed, direct decays to SM quarks are always dominant in the case of the presence of a sizeable mixing with SM quarks, as common in explicit models.
\item \emph{Decays into other states}: we have not included decays such as $Q\to q V_{\mathrm{BSM}}$, with $V_{\mathrm{BSM}}$ a new boson present in the model and $q$ a SM quark, since these EW processes are highly model dependent and also because the typical mass limits on $V_{\mathrm{BSM}}$ are higher than those on XQs, so that these decays are usually not kinematically allowed.
\item \emph{Interference effects}: if more than one XQ is present, the possibility of two XQs decaying into an identical final state can lead to interference effects, that has been discussed in Ref.~\cite{Barducci:2013zaa}.
\item \emph{Loop correction to masses and mixing}: EW corrections to mass and mixing of the XQs can in principle remove or add their degeneracies and change their BRs. However, also these effects are highly model dependent and have not been included in the tool. It is left to the user to check whether they are relevant in the model of interest.
\item \emph{Higher order cross sections}: QCD corrections highly affect the pair production cross section. The effect can be added via a model independent $k$-factor under the assumption that the kinematic is unaffected and therefore we have implemented in XQCAT the possibility to choose between the LO and next-to-leading-order (NLO) cross sections. The QCD-corrected cross sections have been computed with the tool described in Ref.~\cite{Cacciari:2011hy}: in this reference the cross sections have been computed at NLO, supplemented by next-to-next-leading-logarithmic resummation (NLO-NNLL) in QCD. Only the central values of the NLO-NNLL cross sections from Ref.~\cite{Cacciari:2011hy} have been implemented for each XQ mass, and XQCAT does not automatically evaulate uncertainties related to the scale dependence of the cross section. However, the user can set the uncertainty on the signal in the code settings (see \ref{app:howto}), and this uncertainty should include -- among other effects -- such scale 
dependence.
\item \emph{Reproduction of experimental results obtained with different analysis techniques}: due to the fact that the analysis techniques adopted in experimental studies are often more refined than the simple cut-and-count analysis considered in the XQCAT framework, we avoid attempting to validate searches which are more sensitive to and require a more detailed understanding of the CMS detector, where the Delphes approach would become invalid, i.e. not where the bulk of the signal events live or when correlations between variables are considered, as in BDT~\cite{Hocker:2007ht} techniques. Therefore a subset of the experimental bins is used in the determination of the eCL (see Sec.~\ref{sec:eCLdetermination}). The procedure of selection of the relevant bins is described in some detail in the validation procedure (see Sec.~\ref{sec:validationprocedure}) and can also be found in Ref.~\cite{Barducci:2014ila}.
\end{itemize}

\subsection{Validation procedure}
\label{sec:validationprocedure}

We have validated our tool by comparing our results to the experimental data for some specific channels and branching ratios, considering the CMS inclusive search~\cite{Chatrchyan:2013uxa,twikiB2G12015}.
The validation consists of two parts and has the purpose of testing the two main sections of our framework: the limit code that computes the eCLs and the code that extracts the efficiencies. 
While for the case of the SUSY searches we refer, as mentioned, to Ref.~\cite{Buchmueller:2013exa}, we describe here below the essential features of the validation procedure (referring again to Ref.~\cite{Barducci:2014ila} for further details).

\subsubsection{Validation of the limit code}

We have performed this validation comparing the observed and expected mass limits provided by the CMS collaboration with the one computed by our code, in order to find any discrepancy between the statistical method that we use with respect to the one used by the experimentalists. Since the validation of the limit code relies entirely on public experimental data and does not require any simulation on our side, we attribute any discrepancy entirely to a different analysis techniques. We found that with our basic cut-and-count technique and with the eCL method described in Sect.\ref{sec:eCLdetermination} we are not able to reproduce the mass bounds considering the single lepton channels. However, by considering only multi-lepton channels we can reproduce the experimental expected and observed limits on mass bounds with a discrepancy of $-8$\% and $-6$\% respectively (see Tab.1 in Ref.~\cite{Barducci:2014ila}). For this reason only multi-lepton channels have been considered in the implementation of this direct 
search into XQCAT.

\subsubsection{Validation of the efficiencies extraction code} 
\label{sec:effvalidation}

The accuracy of the MC simulation, the correct implementation of selection cuts and the correct reproduction of the true detector effects are the parameters that mostly affects the computation of the efficiencies.

\begin{figure}
\begin{center}
\epsfig{file=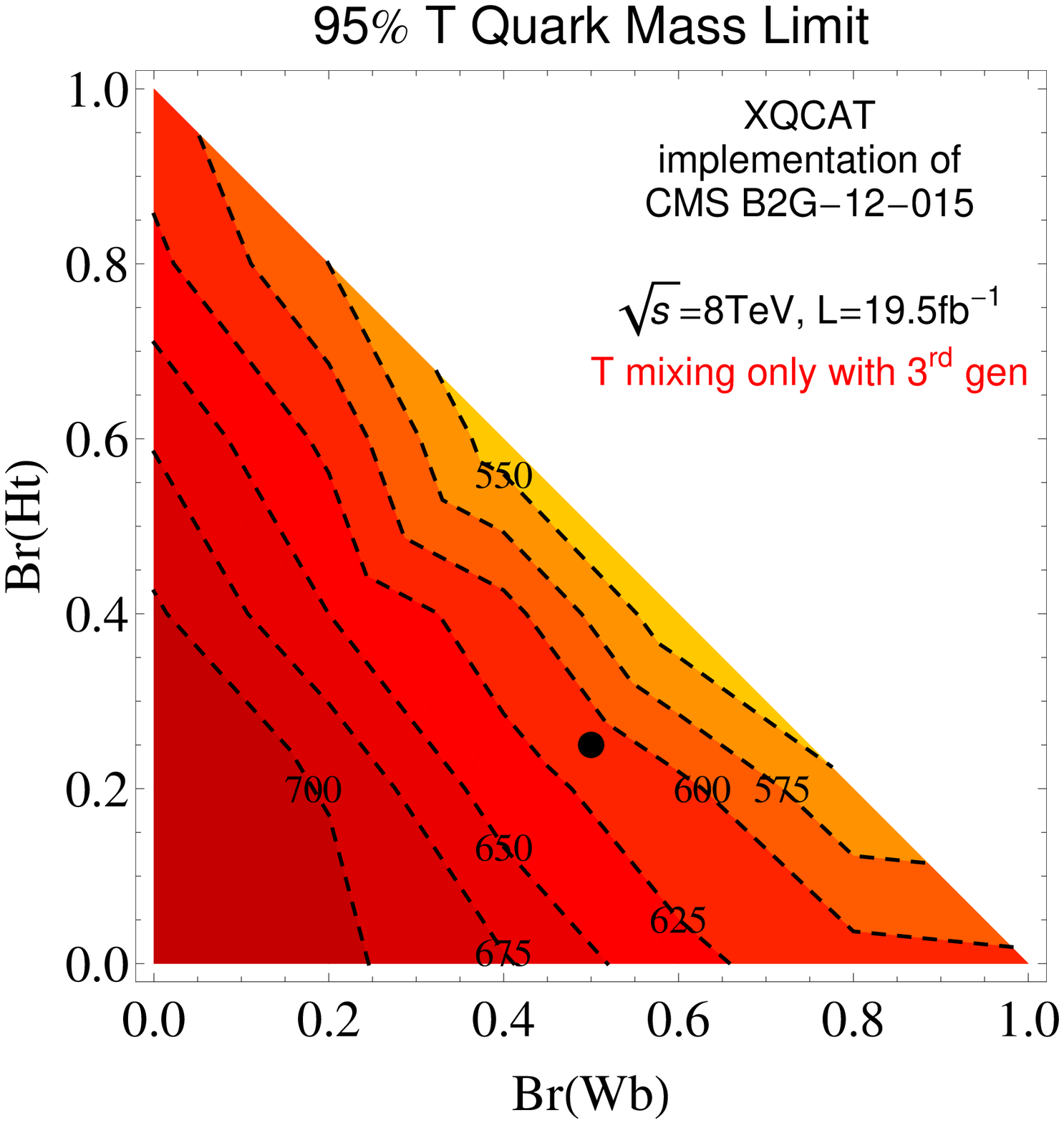,width=.39\textwidth}{(a)}\\
\epsfig{file=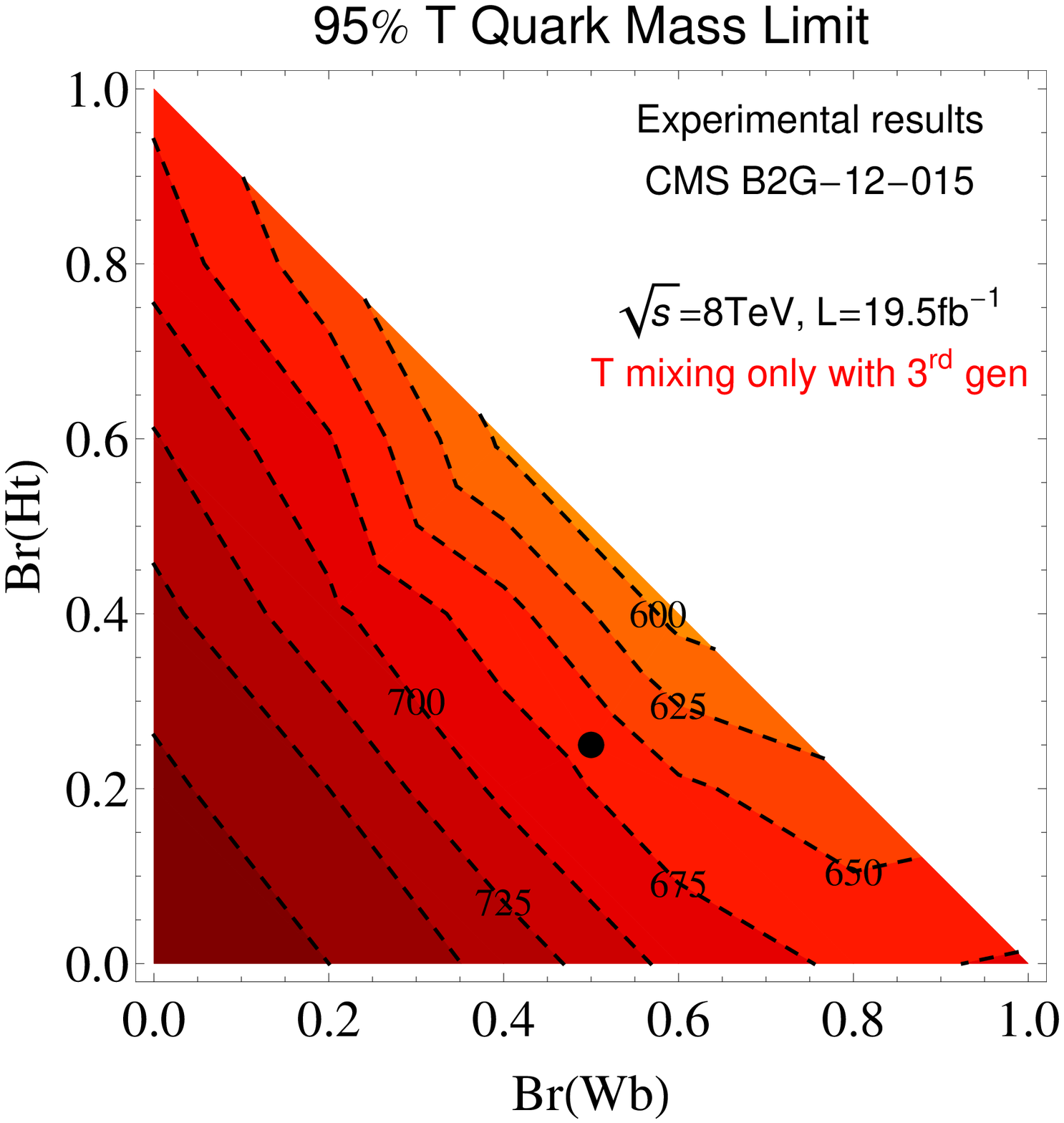,width=.39\textwidth}{(b)}\\
\epsfig{file=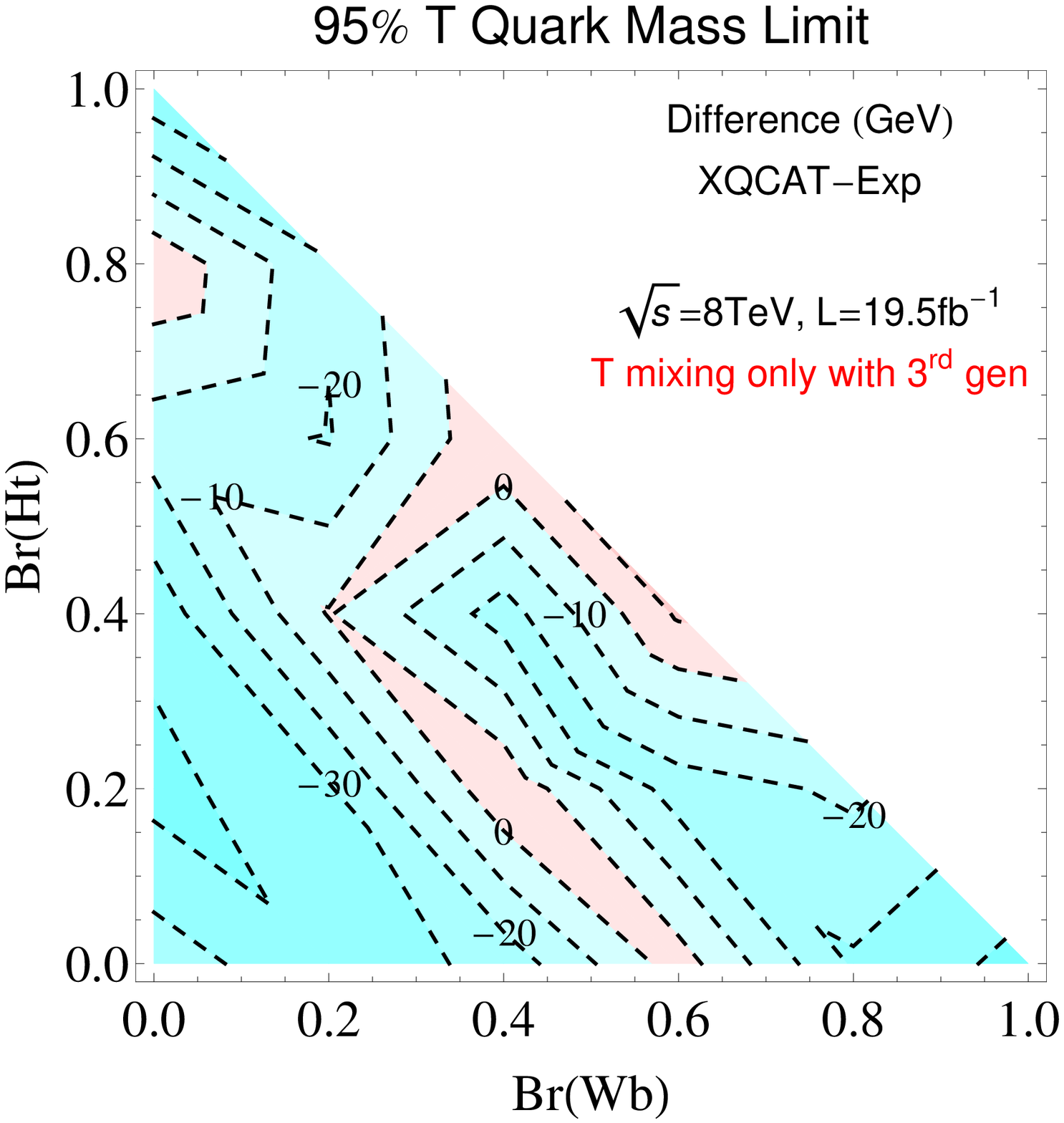,width=.39\textwidth}{(c)}\\
\caption{95\% eCL for a $t^\prime$ in the BR($t^\prime\to Wb$)-BR($t^\prime\to Ht$) plane, for the results obtained with {XQCAT} (a), the experimental results of Refs.~\cite{Chatrchyan:2013uxa,twikiB2G12015} (b) and the difference between the two results (c).}
\label{fig:xqcat-valid}
\end{center}
\end{figure}

\paragraph{Validation of B2G-12-015}
As stated previously in Sect.~\ref{sec:restrictions}, we avoid attempting to validate searches which are more sensitive to and require a more detailed understanding of the CMS detector, where the Delphes approach would become invalid. In the case of B2G-12-015, the single lepton channels are analysed by using BDT techniques, which cannot be reproduced by our cut-and-count approach. Therefore, in the present version of the tool, the calculation of the eCL is performed by considering only three of the multilepton channels. 
In order to validate the implementation of B2G-12-015, we have checked the predictions of our simulation against the results provided by CMS, both by comparing differential distributions for various kinematic observables, and also by comparing the final selection efficiencies for the various analysis channels.
The distribution plots can be found in the webpage of XQCAT~\cite{XQCATwebpage}: in all plots, the solid black line corresponds to the prediction of our simulation while the dashed red line represents the CMS results of Ref.~\cite{twikiB2G12015} for a top partner with a mass of 800 GeV which decays in $W^+ b$, $Zt$ and $Ht$ final states with 50\%, 25\% and 25\% BRs, respectively. The reported CMS results are approximate, as tabulated figures are not provided by the collaboration. Moreover, not all of the plots we have considered appear in the CMS paper: most of the distributions have been published only in Ref.~\cite{twikiB2G12015}. For these reasons, although the distributions we used for validating the analysis are not reported in this paper, to ensure that our results are publicly available we have decided to publish them only on the XQCAT website. 
We find a good agreement in terms of shapes for all distributions, although for some of them the differences in some bins can be up to a factor of a few. As there is no information about how the CMS distributions were obtained, we consider this approximate match an acceptable result. Moreover, the approximate agreement we find at the level of differential distributions is reflected at the level of the selection efficiencies.
Indeed, the discrepancies between the number of signal events computed with our simulation and the ones reported in Ref.~\cite{Chatrchyan:2013uxa,twikiB2G12015} are within $\pm10\%$ for all the multi-lepton channels, except for the second opposite-sign di-lepton, where we find an offset of around -40\% for all masses (see Tab.2 in Ref.~\cite{Barducci:2014ila}).
This difference can however be explained by the differences in implementing the detector effects and/or the selection cuts. We have then decided not to include this channel in our implementation, since a further exploration of these differences would require a more accurate simulation of detector effects, which is not possible with the tools currently available.
As mentioned, we have then restricted the implementation of this direct search to just three channels: OS1, SS and 3L.
Again, the comparison between the efficiencies computed in our simulation and the ones reported in the search webpage~\cite{twikiB2G12015} can be found in the XQCAT webpage~\cite{XQCATwebpage} and not in this paper. 
The validation of XQCAT is finally summarised in Fig.~\ref{fig:xqcat-valid}, where we plot the 95\% eCL for a $t^\prime$ in the BR($t^\prime\to Wb$)-BR($t^\prime\to Ht$) plane, for the results obtained with XQCAT (a), the experimental results of Ref.~\cite{Chatrchyan:2013uxa} (b) and the difference between the two results (c), where we can see that our results are consistent, within a 30 GeV range, for most of the BRs configurations. 

\paragraph{Validation of the SUSY searches} 
The validation of the SUSY searches has been performed in Ref.~\cite{Buchmueller:2013exa} where the authors provided a detailed description of the implementation of these searches. As the very same implementation has been used for the XQCAT simulations, we refer to Ref.~\cite{Buchmueller:2013exa} for more details.

\section{Conclusions and outlook}
\label{sec:conclusions}

The XQCAT framework for the analysis of scenarios with multiple XQs with standard Yukawa couplings to any SM quark and boson ($W$, $Z$ or $H$) has been presented. The code
determines the eCL of any input scenario characterised by masses and BRs of any number of extra quarks with charges $-4/3$, $-1/3$, $+2/3$ and $+5/3$ based on the public data of seven CMS searches at both 7 and 8 TeV. The tool uses a pre-computed database of efficiencies corresponding to the process of QCD-induced pair production and decay of the extra quarks. Details about the simulation procedure and the determination of the eCLs have been extensively provided elsewhere. This paper focused on the computing side of the project and had the purpose of describing the whole XQCAT framework, both the simulation part and the public output, and to be the manual of XQCAT. In contrast, validation and restrictions of the code, as well as techniques for the analysis of scenarios with multiple quarks and interpolation methods, have 
been extensively described in Ref.~\cite{Barducci:2014ila}, as these issues are more about the physical aspects of the analysis than the XQCAT implementation.

As an outlook, we would like to list here future developments of the framework behind the XQCAT code. We are planning to interface our simulation chain with publicly available tools which can perform the efficiency extraction, like CheckMATE and MadAnalysis. These tools contain a large database of implemented searches, and they allow the user to implement (and validate) searches which are not yet present in their database. Due to the cooperative nature of these tools, and to the potential of having a larger database, future versions of XQCAT will be developed exploiting them.
In parallel, we will investigate and validate the emulation of more advanced experimental techniques, e.g. going beyond cut-and-count analyses, considering boosted topologies and so on. We also plan to enlarge the spectrum of signatures and to include potentially relevant effects that can modify the kinematics of the events. Specifically, we aim at the following.
\begin{enumerate}
\item Inclusion of tree-level interference and one-loop mixing effects between XQs, as discussed in Ref.~\cite{Barducci:2013zaa}.
\item Inclusion of processes of single production of XQs.
\item Decays of XQs into invisible states.
\item Systematic treatment of cascade decays of XQs.
\end{enumerate}

Finally, we repeat here for clarity the main webpage of XQCAT (Ref.~\cite{XQCATwebpage}),\\[0.15cm]
\centerline{\tt \href{https://launchpad.net/xqcat}{https://launchpad.net/xqcat}}\\[0.15cm]
\noindent
from where the source files can be found, alongside a log of updates, modifications, bug-fixes, etc... A Q\&A section and a user blog will also be maintained therein.

\appendix

\section{How to use the code}
\label{app:howto}

This Appendix illustrates the general usage of the XQCAT code that will enable a user to start running it
straightforwardly.
The \verb|settings.dat| and \verb|input_default.dat| files present in the \verb|input| folder in the main directory of the code are the unique two files that need to be edited by the user in order to provide the necessary settings and physical input for a run.

The first file, \verb|settings.dat|, presents the following user customizable settings.
\begin{itemize} %DB
\item \verb|iter|: is the number of iterations that the limit code will use to compute the eCL for a given scenario. The default value is set to be 10000. Higher values will guarantee an increase of the stability of the results but the run time will be longer. Conversely, decreasing this value will make the code run faster, but the results will be less stable.

\item \verb|sigunc|: is the uncertainty on the signal rate that will be used into the limit code. The default value of 0.3 takes into account both the uncertainties on the computations of the efficiencies  and the uncertainty on the QCD production cross section.

\item \verb|combination|: if this flag is set to 1 the code will compute the eCLs also for the combinations of the independent, and therefore statistical combinable, searches. If it is set to 0 the eCLs will be computed just for each of the searches independently. The default value is 1.

\item \verb|debug|: if this flag is set to 1 the code will run in a debug mode for which the MC generator that computes the eCLs will use a fixed seed. This allows the user to check the stability of the results with, for example, different choices of \verb|iter| or \verb|sigunc| or also of different choices of physical input parameters. If set to 0 the code will use a random seed for the determination of the eCLs.
\end{itemize}

The \verb|input_default.dat| file is an example of how the physical inputs should be given to XQCAT.
When the code is run it will search for a file called \verb|input.dat| with the desired choice of physical input that, if not found, will be created copying the \verb|input_default.dat| one.
The input card requires the user to provide the basic information of a BSM model with an XQ sector that are masses, BRs and chiralities of the couplings for each of the XQ with a given electric charge that in the card are denoted as \verb|XVLQ|,\verb|TVLQ|,\verb|BVLQ| and \verb|YVLQ| for quarks with charge $+5/3$, $+2/3$,$-1/3$ and $-4/3$, respectively.
The default input card that comes with the code is:
\vskip 8pt

{\footnotesize{
\hskip -7pt\verb|#XVLQ: Mass Wu Wc Wt Chir(L=1, R=2)|

\hskip -7pt\verb|#ENDXVLQ|
\vskip 8pt
\hskip -7pt\verb|#YVLQ: Mass Wd Ws Wb Chir(L=1, R=2)|

\hskip -7pt\verb|#ENDYVLQ|
\vskip 8pt
\hskip -7pt\verb|#TVLQ: Mass Wd Ws Wb Zu Zc Zt Hu Hc Ht Chir(L=1, R=2)|

\hskip -7pt\verb|1.00000000E+03 1 0 0 0 0 0 0 0 0 1|

\hskip -7pt\verb|#ENDTVLQ|
\vskip 8pt
\hskip -7pt\verb|#BVLQ: Mass Wu Wc Wt Zd Zs Zb Hd Hs Hb Chir(L=1, R=2)|

\hskip -7pt\verb|1.00000000E+03 1 0 0 0 0 0 0 0 0 1|

\hskip -7pt\verb|6.00000000E+02 1 0 0 0 0 0 0 0 0 1|

\hskip -7pt\verb|#ENDBVLQ|
}}
\vskip 8pt
\noindent
correspondingly to which XQCAT will calculate the eCLs for a scenario with a quark of charge $+2/3$ that has a 100\% BR into $W^+d$ with a coupling with a dominant left chirality and with two quarks of charge $-1/3$
that have a 100\% BR into $W^-u$ with again a dominant left chirality for the couplings.
The \emph{header} and \emph{footer} that identify each quark species should not be modified by the user and they should be exactly above and below the first and the last row of information on the quarks, with no blank lines left between (that are not allowed either between the various quarks for each specie). Finally, between mass, BR and chirality information entries there should be one space character (no tab characters are allowed in the input card) and a return to new line character after the chirality information.
There is no limit on the number of XQs that can be used, although with a larger spectrum the code will take more time to compute the eCLs.

In order to run the code the user needs to type \verb|./xqcat.pl| into a command line in the main folder of the program after which the presentation picture of XQCAT will tell the user if the code is running or not in debug mode.
XQCAT will ask if the user wants to compute the results using LO, with the choice 1, or NNLO+NNLL with the choice 2, QCD-induced pair production cross section.
After this it will ask for the \verb|user_run| name of the folder in which the run results will be stored, that can then be found in the folder \verb|Results| in the main directory of the program. No spaces are allowed in assigning the results folder name.
If a run with the chosen name is already present in the \verb|Results| directory, XQCAT will ask whether to cancel it, with the choice 1, or rename it, with the choice 2, and in the latter case it will ask for the name to which rename the old run folder.
After this the code will start to run and compute the eCLs for the given scenario printing in stdout mode the search for which is calculating the eCL, the progress on the number of iterations and the final information for the given search (or combination): $-2\log{Q}$ and the final eCL.
During the run the code will also compute the sum of the BRs for each XQ giving an error message if the sum is above 1, after which the code will exit. In the case that the sum of the BRs is below 1 the code will give the user a warning, but it will continue the calculation. This is motivated by the fact that XQCAT can consider just decays into SM quarks and bosons, but in principle there could be extra decay channels present in the user's model that will reduce the rates into SM final states.
The code can also be run in batch mode, in which case the user should type into a command line in the main folder of the program \verb|./xqcat.pl <1 or 2> <run_name>|.
Please notice however that if the folder \verb|run_name| is already present, the code will ask interactively what to do with the old folder even if the code is running in batch mode.
When the code is run in batch mode it is then necessary to assign a non-existing result folder name or cancel the folder after each run if performing multiple runs with the same \verb|run_name| in order to avoid the code to go in interactive mode.
The present version of the code does not allow a parameter scan, the \verb|input.dat| files corresponds to a single configuration in the parameter space.
In case the user wishes to perform a parameter scan it is necessary to write a script which should rename or erase the previous result folder since the code will probably be run in batch mode.
A scan script written in Perl can be provided upon request and will be included in the next version of the code.

The results of the run are summarised in the file \verb|eCLs_summary.dat| in \verb|Results/user_run|
in which the information on the eCLs for each search (and, in case, their combination) are stored.
In the \verb|run_files| subfolder the user can find the files that have been used by XQCAT during the computation, which  are divided in three directories, as follows.
\begin{itemize}
\item \verb|eCLs|: these files, identified by the name of the corresponding search, report the values of $-2\log{Q}$ and eCL for each search (or combination). We chose to keep them since it could be easier to extract the final results from these files than from \verb|eCLs_summary.dat| in  case that XQCAT is run with a script to allow, for example, a parameter scan for a given model.
\item \verb|cpp|: these \verb|.cpp| files, again identified by the corresponding search name, contain the  \verb|ROOT| based code used by XQCAT for the computation of the eCL and the physical information for each bin of the search.
They are generated by the code from the prototype files contained in \verb|core/limit_code| and are filled with the physical information for signal, background and data that are contained in the files created from the prototype in \verb|core/exp_data_proto|. Further details can be found in~\ref{app:structure}.

In particular  \verb|bg_orig[i]|, \verb|bguncert_orig[i]|, \verb|sig_orig[i]|, \verb|sigunc_orig[i]|, \verb|data[i]| 
are the background, background uncertainty, signal, signal uncertainty and data for the $i^{th}$ bin of the search where $i=0,n_{bin}-1$. Except for the quantities related to the signal, that are computed by XQCAT, the other values can be found in the corresponding experimental papers for which 
the associated $arXiv$ number can be found in \verb|eCLs_summary.dat|.
\item \verb|root|: these \verb|.root| files, again identified by the corresponding search name, contain the histograms for the background and signal plus background distributions, together with the confidence eCL histogram.
\end{itemize}

Finally, the \verb|core| folder present in the main directory contains all the subroutines of the code together with the database information that XQCAT uses and that are described  in~\ref{app:structure}.

\section{Structure of the code}
\label{app:structure}

In this Appendix we describe the core of the XQCAT code, which can be found in the \verb|core| folder that contains the \verb|cs_database|, \verb|efficiencies_database|, \verb|exp_data_proto|, \verb|limitcode| and \verb|subroutine| directories where the first three are the physics database of XQCAT.

\subsection{Database information}

The~\verb|cs_database| directory contains four files:\\[0.005cm]

\verb|cs_7TeV.dat|

\verb|cs_8TeV.dat|

\verb|cs_7TeV_NLO_NNLL.dat| 

\verb|cs_8TeV_NLO_NNLL.dat|\\[0.005cm]

\noindent
 which contain the values of the QCD-induced pair productions cross sections at 7 and 8 TeV at LO and NLO+NNLL for an XQ with a mass from 400 GeV to 2000 GeV in steps of 1 GeV.

The \verb|efficiencies_database| directory contains, as described in Sec.~\ref{sec:effdatabase}, the efficiencies information for the 7 and 8 TeV data implemented for all the simulated masses, combinations of decay channels and coupling chiralities.

The \verb|exp_data_proto| directory contains the files with the background, error on the background and observed data information and are named with the corresponding search name.
For example, the file \verb|vlq_CMS7_SS_analysis5_PROTO.txt| in the \verb|7TeV/CMS| subfolder contains the information for the three bins of the 7 TeV CMS search for SS dileptons and the file structure is
\vskip 8pt

{\footnotesize{
\hskip -7pt\verb|1.1 1 SED_1BIN SED_SIGUNC 1|

\hskip -7pt\verb|1.2 1 SED_2BIN SED_SIGUNC 0|

\hskip -7pt\verb|2.6 0.54 SED_3BIN SED_SIGUNC 3|
}}
\vskip8pt
\noindent
where the two strings for each line starting with the \verb|SED_| identifier will be overwritten by the code with the number of signal events and uncertainties on the signal, respectively. When the code runs temporary files called with the corresponding search name, \verb|vlq_CMS7_SS_analysis5.txt| in this case, will be created in a temporary directory \verb|exp_data| and will then be read by the \verb|cl_calculation.pl| subroutine, see~\ref{subsec:subroutines}, to create the \verb|.cpp| limit code files. This temporary directory and files will be removed at the end of the run.
Information on background and data can be found in the experimental papers for which the corresponding $arXiv$ number can be found in \verb|eCLs_summary.dat|

\subsection{Limit code}
\label{subsec:limitcode}

The limit code directory contains the following files:\\

\verb|makefile_PROTO|

\verb|statistics-5_header_PROTO.cpp|

\verb|statistics-5_footer_PROTO.cpp| 

\verb|statistics-5_footer_debug_PROTO.cpp|\\

\noindent
The header and footer are the prototype files that will be used by XQCAT, together with the information contained in the files created from the prototype in \verb|core/exp_data_proto|, to generate the \verb|.cpp| files containing all the physical information for each bin of the search or combination and that can be found at the end of the run in the \verb|Results/user_run/cpp| directory.
As explained, these are the files with which XQCAT will compute the eCLs for the given search and combinations. 
In debug mode the footer labelled with \verb|_debug| will be used in order to force the limit code to work with a fixed random seed.
Also the \verb|makefile| will be created from the prototype, though in this version of the code no customization is available.

\subsection{Subroutines}
\label{subsec:subroutines}

XQCAT uses seven Perl subroutines in order to calculate the eCLs from the user input database 

\begin{itemize}
\item \verb|importdata()|: this function reads the information in the input card created by the user or from the default file. In particular, the number of quarks for each specie, masses, BRs and chirality indices will be saved in arrays used to calculate the final number of signal events.
\item \verb|findlimits(massmin,massmax)|: this function compares the minimum and maximum value of the quarks masses given in input by the user with the simulated masses present in the database. If \verb|massmin| is lower or \verb|massmax| is higher than the highest simulated mass, the code will give an error and stop.
\item \verb|findeff(energy,detector,mass,vlqname,| 
\verb|vlqdecay1,vlqdecay2,chirality,search)|: this function calculates the efficiencies for a given LHC \verb|energy| and \verb|detector| \verb|search| for a quark of type \verb|vlqname| with a given \verb|mass| and \verb|chirality| for the final state given by the quark and antiquark decays \verb|vlqdecay1| and \verb|vlqdecay2|.
The function will search in the efficiency database for the \verb|mass| of the given quark. If this file is not present a linear interpolation between the closest lower and higher simulated masses will be computed.
\item \verb|findcs(energy,mass,csinput)|: this function calculates the cross section for the QCD production of a pair of quarks for a given LHC \verb|energy| and quark \verb|mass|. LO or NNLO+NNLL cross section will be computed according to the value assigned to \verb|csinput|, respectively 1 and 2.
If the quark mass is not present in the database a linear interpolation between the closest lower and higher cross sections in the database will be computed.
\item \verb|exportdata(ScriptPath,energy,detector,|\\
\verb|search,nbin,weightvlqtypesumeff)|: this function will create a temporary file from the proto files present \verb|exp_data_proto|, as described in the previous subsection, for each of the LHC \verb|energy|, \verb|detector| and \verb|searches|, or combination, with a corresponding number of bins, \verb|nbins|, for which the code runs and in which it will be printed the final value of the signal rate that is contained in the vector \verb|weightvlqtypesumeff|.
\item \verb|cl_calculation(energy,detector,search,|\\
\verb|searchname)|: this function creates and runs the \verb|.cpp| files with which the code computes the eCLs for each of the LHC \verb|energy|, \verb|detector| and \verb|searches| for which the code run. An user friendly name of the search, \verb|searchname|, is also required to be printed in stdout and in \verb|eCLs_summary.dat|. 
\item \verb|cl_combination(ENERGY,DETECTOR,SEARCH,|\\
\verb|SEARCHNAME)|: this function runs the combination of searches and, analogously to \verb|cl_calculation|, it runs the \verb|.cpp| files with which the code computes the eCLs for each of the LHC \verb|ENERGY|, \verb|DETECTOR| and the list of \verb|SEARCHES| for which the code run. The list of correlated searches is hardcoded in the current version, but the user can modify the file to perform combinations on different sets; notice however that combinations require that both signal and background bins are not correlated.

\end{itemize}

\section{Worked out example}
\label{app:example}

Although the use of XQCAT is relatively straightforward, in this final appendix, we provide two worked out examples and the corresponding results that have been calculated in debug mode, with 10000 iterations, an uncertainty on the signal of 0.3 and calculating also the results for the combinations of the searches.

\subsection{First example}

As a first example we do not create a \verb|input.dat| card but we use the one that comes by default with XQCAT.
Typing at terminal \verb|./xqcat.pl| the user will find the following stdout
\vskip 8pt
{\footnotesize{
\hskip -7pt\verb|---------------------------------|

\hskip -7pt\verb|---                           ---|

\hskip -7pt\verb|------                     ------|

\hskip -7pt\verb|--------    xqcat_v1.2   --------|

\hskip -7pt\verb|---                           ---|

\hskip -7pt\verb|----- WARNING: debug mode!! -----|

\hskip -7pt\verb|--- to run it in default mode ---|

\hskip -7pt\verb|---- set the debug flag to 0 ----|

\hskip -7pt\verb|----     in settings.dat      ---|

\hskip -7pt\verb|------                     ------|

\hskip -7pt\verb|---                           ---|

\hskip -7pt\verb|---------------------------------|

\hskip -7pt\verb|Please select 1 if you want results|\\
\hskip -7pt\verb|at LO or 2 if you want results at NLO+NNLL|
}}
\vskip 8pt

\noindent

\noindent
In making the choice \verb|2| the user will be asked to give a name for the run
\vskip 8pt
{\footnotesize{
\hskip -7pt\verb|Please type the name of the run|
}}
\vskip 8pt

\noindent
that we call \verb|test_run_1|. Since no previous results are present the code will start running and printing in stdout mode 
the eCLs for the searches as soon as they are calculated
\vskip 8pt
{\footnotesize{
\hskip -7pt\verb|input.dat not present using the default one|
}}
\vskip 8pt
{\footnotesize{
\hskip -7pt\verb|Calculating exlusion CL for|

\hskip -7pt\verb|CMS alphaT 7TeV (arXiv:1210.8115)|

\hskip -7pt\verb|Iteration 10000 out of 10000|

\hskip -7pt\verb|-2*ln(Q) = 0.584964|

\hskip -7pt\verb|eCL = 0.2947541|
}}
\vskip 8pt

Once all the eCLs are calculated the program will end
\vskip 8pt
{\footnotesize{
\hskip -7pt\verb|---------------------------------|

\hskip -7pt\verb|--- Thank you for using xqcat! --|

\hskip -7pt\verb|---------------------------------|
}}

\vskip 8pt
\noindent
and the user will find the results in the \verb|Results/test_run_1| folder.

\subsection{Second example}

After having run the first example a card called \verb|input.dat| will now be present in the \verb|input| folder that can be modified by the user before running again XQCAT.
We now enter in stdin mode \verb|./xqcat.pl 1 test_run_1| where the argument for the choice of the QCD cross section, LO in this case, and the name of the run are given to XQCAT directly from the command line.
Since the run name is the same as the one of the first example, XQCAT will ask the user what to do with the old results.
\vskip 8pt
{\footnotesize{
\hskip -7pt\verb|Result test_run_1 already exixts:|\\
\hskip -7pt\verb|select 1 if you want to remove it|\\
\hskip -7pt\verb|or 2 if you want to rename it|
}}
\vskip 8pt

\noindent
By choosing \verb|2| the code will ask a new name to be assigned to the old folder
\vskip 8pt
{\footnotesize{
\hskip -7pt\verb|Type the name of the folder to which you|\\
\hskip -7pt\verb|wish to rename test_run_1 to|
}}
\vskip 8pt

\noindent
that we call \verb|test_run_1_old|.
The code will start running as in the previous example giving the corresponding LO results.
\vskip 8pt
{\footnotesize{
\hskip -7pt\verb|Calculating exlusion CL for|

\hskip -7pt\verb|CMS alphaT 7TeV (arXiv:1210.8115)|

\hskip -7pt\verb|Iteration 10000 out of 10000|

\hskip -7pt\verb|-2*ln(Q) = 0.37357|

\hskip -7pt\verb|eCL = 0.19685039|
}}
\vskip 8pt
\noindent
and the user will finally find both the new and old (renamed) results in the \verb|Results| subfolder.

\noindent

%% References
%% Usage of~\cite is as follows:
%%  ~\cite{key}         ==>>  [#]
%%  ~\cite[chap. 2]{key} ==>> [#, chap. 2]
%%


% \section*{References}
% \bibliographystyle{elsarticle-num}
\bibliographystyle{JHEP}
\bibliography{XQCAT}

\providecommand{\href}[2]{#2}\begingroup\raggedright\begin{thebibliography}{10}

\bibitem{Dobrescu:1997nm}
B.~A. Dobrescu and C.~T. Hill, {\it {Electroweak symmetry breaking via top
  condensation seesaw}},  {\em Phys.Rev.Lett.} {\bf 81} (1998) 2634--2637,
  [\href{http://xxx.lanl.gov/abs/hep-ph/9712319}{{\tt hep-ph/9712319}}].

\bibitem{Chivukula:1998wd}
R.~S. Chivukula, B.~A. Dobrescu, H.~Georgi, and C.~T. Hill, {\it {Top quark
  seesaw theory of electroweak symmetry breaking}},  {\em Phys.Rev.} {\bf D59}
  (1999) 075003, [\href{http://xxx.lanl.gov/abs/hep-ph/9809470}{{\tt
  hep-ph/9809470}}].

\bibitem{He:2001fz}
H.-J. He, C.~T. Hill, and T.~M. Tait, {\it {Top quark seesaw, vacuum structure
  and electroweak precision constraints}},  {\em Phys.Rev.} {\bf D65} (2002)
  055006, [\href{http://xxx.lanl.gov/abs/hep-ph/0108041}{{\tt
  hep-ph/0108041}}].

\bibitem{Hill:2002ap}
C.~T. Hill and E.~H. Simmons, {\it {Strong dynamics and electroweak symmetry
  breaking}},  {\em Phys.Rept.} {\bf 381} (2003) 235--402,
  [\href{http://xxx.lanl.gov/abs/hep-ph/0203079}{{\tt hep-ph/0203079}}].

\bibitem{Agashe:2004rs}
K.~Agashe, R.~Contino, and A.~Pomarol, {\it {The Minimal composite Higgs
  model}},  {\em Nucl.Phys.} {\bf B719} (2005) 165--187,
  [\href{http://xxx.lanl.gov/abs/hep-ph/0412089}{{\tt hep-ph/0412089}}].

\bibitem{Contino:2006qr}
R.~Contino, L.~Da~Rold, and A.~Pomarol, {\it {Light custodians in natural
  composite Higgs models}},  {\em Phys.Rev.} {\bf D75} (2007) 055014,
  [\href{http://xxx.lanl.gov/abs/hep-ph/0612048}{{\tt hep-ph/0612048}}].

\bibitem{Barbieri:2007bh}
R.~Barbieri, B.~Bellazzini, V.~S. Rychkov, and A.~Varagnolo, {\it {The Higgs
  boson from an extended symmetry}},  {\em Phys.Rev.} {\bf D76} (2007) 115008,
  [\href{http://xxx.lanl.gov/abs/0706.0432}{{\tt arXiv:0706.0432}}].

\bibitem{Anastasiou:2009rv}
C.~Anastasiou, E.~Furlan, and J.~Santiago, {\it {Realistic Composite Higgs
  Models}},  {\em Phys.Rev.} {\bf D79} (2009) 075003,
  [\href{http://xxx.lanl.gov/abs/0901.2117}{{\tt arXiv:0901.2117}}].

\bibitem{Gripaios:2014pqa}
B.~Gripaios, T.~Mueller, M.~Parker, and D.~Sutherland, {\it {Search Strategies
  for Top Partners in Composite Higgs models}},
  \href{http://xxx.lanl.gov/abs/1406.5957}{{\tt arXiv:1406.5957}}.

\bibitem{ArkaniHamed:2002qx}
N.~Arkani-Hamed, A.~Cohen, E.~Katz, A.~Nelson, T.~Gregoire, et~al., {\it {The
  Minimal moose for a little Higgs}},  {\em JHEP} {\bf 0208} (2002) 021,
  [\href{http://xxx.lanl.gov/abs/hep-ph/0206020}{{\tt hep-ph/0206020}}].

\bibitem{Yang:2014mba}
B.~Yang, G.~Mi, and N.~Liu, {\it {Higgs couplings and Naturalness in the
  littlest Higgs model with T-parity at the LHC and TLEP}},
  \href{http://xxx.lanl.gov/abs/1407.6123}{{\tt arXiv:1407.6123}}.

\bibitem{Rosner:1985hx}
J.~L. Rosner, {\it {E6 and Exotic Fermions}},  {\em Comments Nucl.Part.Phys.}
  {\bf 15} (1986) 195.

\bibitem{Robinett:1985dz}
R.~Robinett, {\it {On the Mixing and Production of Exotic Fermions in E6}},
  {\em Phys.Rev.} {\bf D33} (1986) 1908.

\bibitem{Antoniadis:1990ew}
I.~Antoniadis, {\it {A Possible new dimension at a few TeV}},  {\em Phys.Lett.}
  {\bf B246} (1990) 377--384.

\bibitem{Appelquist:2000nn}
T.~Appelquist, H.-C. Cheng, and B.~A. Dobrescu, {\it {Bounds on universal extra
  dimensions}},  {\em Phys.Rev.} {\bf D64} (2001) 035002,
  [\href{http://xxx.lanl.gov/abs/hep-ph/0012100}{{\tt hep-ph/0012100}}].

\bibitem{Csaki:2003sh}
C.~Csaki, C.~Grojean, J.~Hubisz, Y.~Shirman, and J.~Terning, {\it {Fermions on
  an interval: Quark and lepton masses without a Higgs}},  {\em Phys.Rev.} {\bf
  D70} (2004) 015012, [\href{http://xxx.lanl.gov/abs/hep-ph/0310355}{{\tt
  hep-ph/0310355}}].

\bibitem{Cacciapaglia:2009pa}
G.~Cacciapaglia, A.~Deandrea, and J.~Llodra-Perez, {\it {A Dark Matter
  candidate from Lorentz Invariance in 6D}},  {\em JHEP} {\bf 1003} (2010) 083,
  [\href{http://xxx.lanl.gov/abs/0907.4993}{{\tt arXiv:0907.4993}}].

\bibitem{twikiATLAS}
\url{https://twiki.cern.ch/twiki/bin/view/AtlasPublic/ExoticsPublicResults}.

\bibitem{twikiCMS}
\url{https://twiki.cern.ch/twiki/bin/view/CMSPublic/PhysicsResultsB2G}.

\bibitem{delAguila:2000rc}
F.~del Aguila, M.~Perez-Victoria, and J.~Santiago, {\it {Observable
  contributions of new exotic quarks to quark mixing}},  {\em JHEP} {\bf 0009}
  (2000) 011, [\href{http://xxx.lanl.gov/abs/hep-ph/0007316}{{\tt
  hep-ph/0007316}}].

\bibitem{Cacciapaglia:2012dd}
G.~Cacciapaglia, A.~Deandrea, L.~Panizzi, S.~Perries, and V.~Sordini, {\it
  {Heavy Vector-like quark with charge 5/3 at the LHC}},  {\em JHEP} {\bf 1303}
  (2013) 004, [\href{http://xxx.lanl.gov/abs/1211.4034}{{\tt
  arXiv:1211.4034}}].

\bibitem{Okada:2012gy}
Y.~Okada and L.~Panizzi, {\it {LHC signatures of vector-like quarks}},  {\em
  Adv.High Energy Phys.} {\bf 2013} (2013) 364936,
  [\href{http://xxx.lanl.gov/abs/1207.5607}{{\tt arXiv:1207.5607}}].

\bibitem{Read:2000ru}
A.~L. Read, {\it {Modified frequentist analysis of search results (The CL(s)
  method)}},  {\em {CERN-OPEN-2000-205}} (2000).

\bibitem{Read:2002hq}
A.~L. Read, {\it {Presentation of search results: The CL(s) technique}},  {\em
  J.Phys.} {\bf G28} (2002) 2693--2704.

\bibitem{Root}
\url{http://root.cern.ch/drupal/}.

\bibitem{Kraml:2013mwa}
S.~Kraml, S.~Kulkarni, U.~Laa, A.~Lessa, W.~Magerl, et~al., {\it {SModelS: a
  tool for interpreting simplified-model results from the LHC and its
  application to supersymmetry}},  {\em Eur.Phys.J.} {\bf C74} (2014) 2868,
  [\href{http://xxx.lanl.gov/abs/1312.4175}{{\tt arXiv:1312.4175}}].

\bibitem{Kraml:2014sna}
S.~Kraml, S.~Kulkarni, U.~Laa, A.~Lessa, V.~Magerl, et~al., {\it {SModelS v1.0:
  a short user guide}},  \href{http://xxx.lanl.gov/abs/1412.1745}{{\tt
  arXiv:1412.1745}}.

\bibitem{Papucci:2014rja}
M.~Papucci, K.~Sakurai, A.~Weiler, and L.~Zeune, {\it {Fastlim: a fast LHC
  limit calculator}},  {\em Eur.Phys.J.} {\bf C74} (2014), no.~11 3163,
  [\href{http://xxx.lanl.gov/abs/1402.0492}{{\tt arXiv:1402.0492}}].

\bibitem{Drees:2013wra}
M.~Drees, H.~Dreiner, D.~Schmeier, J.~Tattersall, and J.~S. Kim, {\it
  {CheckMATE: Confronting your Favourite New Physics Model with LHC Data}},
  {\em Comput.Phys.Commun.} {\bf 187} (2014) 227--265,
  [\href{http://xxx.lanl.gov/abs/1312.2591}{{\tt arXiv:1312.2591}}].

\bibitem{Kim:2015wza}
J.~S. Kim, D.~Schmeier, J.~Tattersall, and K.~Rolbiecki, {\it {A framework to
  create customised LHC analyses within CheckMATE}},
  \href{http://xxx.lanl.gov/abs/1503.0112}{{\tt arXiv:1503.0112}}.

\bibitem{Conte:2012fm}
E.~Conte, B.~Fuks, and G.~Serret, {\it {MadAnalysis 5, A User-Friendly
  Framework for Collider Phenomenology}},  {\em Comput.Phys.Commun.} {\bf 184}
  (2013) 222--256, [\href{http://xxx.lanl.gov/abs/1206.1599}{{\tt
  arXiv:1206.1599}}].

\bibitem{Dumont:2014tja}
B.~Dumont, B.~Fuks, S.~Kraml, S.~Bein, G.~Chalons, et~al., {\it {Toward a
  public analysis database for LHC new physics searches using MADANALYSIS 5}},
  {\em Eur.Phys.J.} {\bf C75} (2015), no.~2 56,
  [\href{http://xxx.lanl.gov/abs/1407.3278}{{\tt arXiv:1407.3278}}].

\bibitem{Conte:2014zja}
E.~Conte, B.~Dumont, B.~Fuks, and C.~Wymant, {\it {Designing and recasting LHC
  analyses with MadAnalysis 5}},  {\em Eur.Phys.J.} {\bf C74} (2014), no.~10
  3103, [\href{http://xxx.lanl.gov/abs/1405.3982}{{\tt arXiv:1405.3982}}].

\bibitem{Barducci:2014ila}
D.~Barducci, A.~Belyaev, M.~Buchkremer, G.~Cacciapaglia, A.~Deandrea, et~al.,
  {\it {Model Independent Framework for Analysis of Scenarios with Multiple
  Heavy Extra Quarks}},  \href{http://xxx.lanl.gov/abs/1405.0737}{{\tt
  arXiv:1405.0737}}.

\bibitem{Buchmueller:2013exa}
O.~Buchmueller and J.~Marrouche, {\it {Universal mass limits on gluino and
  third-generation squarks in the context of Natural-like SUSY spectra}},  {\em
  Int.J.Mod.Phys.} {\bf A29} (2014) 1450032,
  [\href{http://xxx.lanl.gov/abs/1304.2185}{{\tt arXiv:1304.2185}}].

\bibitem{XQCATwebpage}
\url{https://launchpad.net/xqcat}.

\bibitem{delAguila:1982fs}
F.~del Aguila and M.~J. Bowick, {\it {The Possibility of New Fermions With
  $\Delta$ I = 0 Mass}},  {\em Nucl.Phys.} {\bf B224} (1983) 107.

\bibitem{AguilarSaavedra:2009es}
J.~Aguilar-Saavedra, {\it {Identifying top partners at LHC}},  {\em JHEP} {\bf
  0911} (2009) 030, [\href{http://xxx.lanl.gov/abs/0907.3155}{{\tt
  arXiv:0907.3155}}].

\bibitem{DeSimone:2012fs}
A.~De~Simone, O.~Matsedonskyi, R.~Rattazzi, and A.~Wulzer, {\it {A First Top
  Partner Hunter's Guide}},  {\em JHEP} {\bf 1304} (2013) 004,
  [\href{http://xxx.lanl.gov/abs/1211.5663}{{\tt arXiv:1211.5663}}].

\bibitem{Buchkremer:2013bha}
M.~Buchkremer, G.~Cacciapaglia, A.~Deandrea, and L.~Panizzi, {\it {Model
  Independent Framework for Searches of Top Partners}},  {\em Nucl.Phys.} {\bf
  B876} (2013) 376--417, [\href{http://xxx.lanl.gov/abs/1305.4172}{{\tt
  arXiv:1305.4172}}].

\bibitem{Alwall:2014hca}
J.~Alwall, R.~Frederix, S.~Frixione, V.~Hirschi, F.~Maltoni, et~al., {\it {The
  automated computation of tree-level and next-to-leading order differential
  cross sections, and their matching to parton shower simulations}},  {\em
  JHEP} {\bf 1407} (2014) 079, [\href{http://xxx.lanl.gov/abs/1405.0301}{{\tt
  arXiv:1405.0301}}].

\bibitem{Degrande:2011ua}
C.~Degrande, C.~Duhr, B.~Fuks, D.~Grellscheid, O.~Mattelaer, et~al., {\it {UFO
  - The Universal FeynRules Output}},  {\em Comput.Phys.Commun.} {\bf 183}
  (2012) 1201--1214, [\href{http://xxx.lanl.gov/abs/1108.2040}{{\tt
  arXiv:1108.2040}}].

\bibitem{feynrulesVLQ}
\url{http://feynrules.irmp.ucl.ac.be/wiki/VLQ}.

\bibitem{Christensen:2008py}
N.~D. Christensen and C.~Duhr, {\it {FeynRules - Feynman rules made easy}},
  {\em Comput.Phys.Commun.} {\bf 180} (2009) 1614--1641,
  [\href{http://xxx.lanl.gov/abs/0806.4194}{{\tt arXiv:0806.4194}}].

\bibitem{Alloul:2013bka}
A.~Alloul, N.~D. Christensen, C.~Degrande, C.~Duhr, and B.~Fuks, {\it
  {FeynRules 2.0 - A complete toolbox for tree-level phenomenology}},  {\em
  Comput.Phys.Commun.} {\bf 185} (2014) 2250--2300,
  [\href{http://xxx.lanl.gov/abs/1310.1921}{{\tt arXiv:1310.1921}}].

\bibitem{Pumplin:2002vw}
J.~Pumplin, D.~Stump, J.~Huston, H.~Lai, P.~M. Nadolsky, et~al., {\it {New
  generation of parton distributions with uncertainties from global QCD
  analysis}},  {\em JHEP} {\bf 0207} (2002) 012,
  [\href{http://xxx.lanl.gov/abs/hep-ph/0201195}{{\tt hep-ph/0201195}}].

\bibitem{MGscales}
\url{https://cp3.irmp.ucl.ac.be/projects/madgraph/wiki/FAQ-General-13}.

\bibitem{Meade:2007js}
P.~Meade and M.~Reece, {\it {BRIDGE: Branching ratio inquiry / decay generated
  events}},  \href{http://xxx.lanl.gov/abs/hep-ph/0703031}{{\tt
  hep-ph/0703031}}.

\bibitem{Sjostrand:2006za}
T.~Sjostrand, S.~Mrenna, and P.~Z. Skands, {\it {PYTHIA 6.4 Physics and
  Manual}},  {\em JHEP} {\bf 0605} (2006) 026,
  [\href{http://xxx.lanl.gov/abs/hep-ph/0603175}{{\tt hep-ph/0603175}}].

\bibitem{Ovyn:2009tx}
S.~Ovyn, X.~Rouby, and V.~Lemaitre, {\it {DELPHES, a framework for fast
  simulation of a generic collider experiment}},
  \href{http://xxx.lanl.gov/abs/0903.2225}{{\tt arXiv:0903.2225}}.

\bibitem{deFavereau:2013fsa}
{\bf DELPHES 3} Collaboration, J.~de~Favereau et~al., {\it {DELPHES 3, A
  modular framework for fast simulation of a generic collider experiment}},
  {\em JHEP} {\bf 1402} (2014) 057,
  [\href{http://xxx.lanl.gov/abs/1307.6346}{{\tt arXiv:1307.6346}}].

\bibitem{Alwall:2007fs}
J.~Alwall, S.~Hoche, F.~Krauss, N.~Lavesson, L.~Lonnblad, et~al., {\it
  {Comparative study of various algorithms for the merging of parton showers
  and matrix elements in hadronic collisions}},  {\em Eur.Phys.J.} {\bf C53}
  (2008) 473--500, [\href{http://xxx.lanl.gov/abs/0706.2569}{{\tt
  arXiv:0706.2569}}].

\bibitem{Alwall:2008qv}
J.~Alwall, S.~de~Visscher, and F.~Maltoni, {\it {QCD radiation in the
  production of heavy colored particles at the LHC}},  {\em JHEP} {\bf 0902}
  (2009) 017, [\href{http://xxx.lanl.gov/abs/0810.5350}{{\tt
  arXiv:0810.5350}}].

\bibitem{MGmatch1}
\url{https://cp3.irmp.ucl.ac.be/projects/madgraph/wiki/IntroMatching}.

\bibitem{MGmatch2}
\url{https://cp3.irmp.ucl.ac.be/projects/madgraph/wiki/Matching}.

\bibitem{Mangano:2006rw}
M.~L. Mangano, M.~Moretti, F.~Piccinini, and M.~Treccani, {\it {Matching matrix
  elements and shower evolution for top-quark production in hadronic
  collisions}},  {\em JHEP} {\bf 0701} (2007) 013,
  [\href{http://xxx.lanl.gov/abs/hep-ph/0611129}{{\tt hep-ph/0611129}}].

\bibitem{Brown:1991hx}
N.~Brown and W.~J. Stirling, {\it {Finding jets and summing soft gluons: A New
  algorithm}},  {\em Z.Phys.} {\bf C53} (1992) 629--636.

\bibitem{Catani:1993hr}
S.~Catani, Y.~L. Dokshitzer, M.~Seymour, and B.~Webber, {\it {Longitudinally
  invariant $K_t$ clustering algorithms for hadron hadron collisions}},  {\em
  Nucl.Phys.} {\bf B406} (1993) 187--224.

\bibitem{Chatrchyan:2013uxa}
{\bf \textnormal{CMS}} Collaboration, S.~Chatrchyan et~al., {\it {Inclusive
  search for a vector-like T quark with charge 2/3 in pp collisions at
  $\sqrt{s}$ = 8 TeV}},  {\em Phys.Lett.} {\bf B729} (2014) 149--171,
  [\href{http://xxx.lanl.gov/abs/1311.7667}{{\tt arXiv:1311.7667}}].

\bibitem{twikiB2G12015}
\url{https://twiki.cern.ch/twiki/bin/view/CMSPublic/PublicResultsB2G12015AdditionalPlots}.

\bibitem{Chatrchyan:2012wa}
{\bf \textnormal{CMS}} Collaboration, S.~Chatrchyan et~al., {\it {Search for
  supersymmetry in final states with missing transverse energy and 0, 1, 2, or
  at least 3 b-quark jets in 7 TeV pp collisions using the variable alphaT}},
  {\em JHEP} {\bf 1301} (2013) 077,
  [\href{http://xxx.lanl.gov/abs/1210.8115}{{\tt arXiv:1210.8115}}].

\bibitem{Chatrchyan:2012sca}
{\bf \textnormal{CMS}} Collaboration, S.~Chatrchyan et~al., {\it {Search for
  supersymmetry in final states with a single lepton, $b$-quark jets, and
  missing transverse energy in proton-proton collisions at $\sqrt{s}=7$ TeV}},
  {\em Phys.Rev.} {\bf D87} (2013), no.~5 052006,
  [\href{http://xxx.lanl.gov/abs/1211.3143}{{\tt arXiv:1211.3143}}].

\bibitem{Chatrchyan:2012te}
{\bf \textnormal{CMS}} Collaboration, S.~Chatrchyan et~al., {\it {Search for
  new physics in events with opposite-sign leptons, jets, and missing
  transverse energy in $pp$ collisions at $\sqrt{s}=7$ TeV}},  {\em Phys.Lett.}
  {\bf B718} (2013) 815--840, [\href{http://xxx.lanl.gov/abs/1206.3949}{{\tt
  arXiv:1206.3949}}].

\bibitem{Chatrchyan:2012sa}
{\bf \textnormal{CMS}} Collaboration, S.~Chatrchyan et~al., {\it {Search for
  new physics in events with same-sign dileptons and $b$-tagged jets in $pp$
  collisions at $\sqrt{s}=7$ TeV}},  {\em JHEP} {\bf 1208} (2012) 110,
  [\href{http://xxx.lanl.gov/abs/1205.3933}{{\tt arXiv:1205.3933}}].

\bibitem{Chatrchyan:2013lya}
{\bf \textnormal{CMS}} Collaboration, S.~Chatrchyan et~al., {\it {Search for
  supersymmetry in hadronic final states with missing transverse energy using
  the variables AlphaT and b-quark multiplicity in pp collisions at 8 TeV}},
  {\em Eur.Phys.J.} {\bf C73} (2013) 2568,
  [\href{http://xxx.lanl.gov/abs/1303.2985}{{\tt arXiv:1303.2985}}].

\bibitem{Chatrchyan:2012paa}
{\bf \textnormal{CMS}} Collaboration, S.~Chatrchyan et~al., {\it {Search for
  new physics in events with same-sign dileptons and $b$ jets in $pp$
  collisions at $\sqrt{s}=8$ TeV}},  {\em JHEP} {\bf 1303} (2013) 037,
  [\href{http://xxx.lanl.gov/abs/1212.6194}{{\tt arXiv:1212.6194}}].

\bibitem{Barducci:2013zaa}
D.~Barducci, A.~Belyaev, J.~Blamey, S.~Moretti, L.~Panizzi, et~al., {\it
  {Towards model-independent approach to the analysis of interference effects
  in pair production of new heavy quarks}},  {\em JHEP} {\bf 1407} (2014) 142,
  [\href{http://xxx.lanl.gov/abs/1311.3977}{{\tt arXiv:1311.3977}}].

\bibitem{Cacciari:2011hy}
M.~Cacciari, M.~Czakon, M.~Mangano, A.~Mitov, and P.~Nason, {\it {Top-pair
  production at hadron colliders with next-to-next-to-leading logarithmic
  soft-gluon resummation}},  {\em Phys.Lett.} {\bf B710} (2012) 612--622,
  [\href{http://xxx.lanl.gov/abs/1111.5869}{{\tt arXiv:1111.5869}}].

\bibitem{Hocker:2007ht}
A.~Hocker, J.~Stelzer, F.~Tegenfeldt, H.~Voss, K.~Voss, et~al., {\it {TMVA -
  Toolkit for Multivariate Data Analysis}},  {\em PoS} {\bf ACAT} (2007) 040,
  [\href{http://xxx.lanl.gov/abs/physics/0703039}{{\tt physics/0703039}}].

\end{thebibliography}\endgroup


\begin{thebibliography}{99}
% \bibitem{1}Reference 1         % This list should only contain those items referenced in the                 
% \bibitem{2}Reference 2         % Program Summary section.   
% \bibitem{3}Reference 3         % Type references in text as [1], [2], etc.
                                 % This list is different from the bibliography at the end of 
                                 % the Long Write-Up.
\bibitem{1} CMS Collaboration, S. Chatrchyan et al., \textit{Search for supersymmetry in final states with missing transverse energy and 0, 1, 2, or at least 3 b-quark jets in 7 TeV pp collisions using the variable $\alpha_T$}, \textit{JHEP} \textbf{1301} (2013) 077, [arXiv:1210.8115].
\bibitem{2} CMS Collaboration, S. Chatrchyan et al., \textit{Search for supersymmetry in final states with a single lepton, b-quark jets, and missing transverse energy in proton-proton collisions at $\sqrt{s}$ = 7 TeV}, \textit{Phys.Rev.} \textbf{D87} (2013), no. 5 052006, [arXiv:1211.3143].
\bibitem{3}  CMS Collaboration, S. Chatrchyan et al., \textit{Search for new physics in events with opposite-sign leptons, jets, and missing transverse energy in $pp$ collisions at $\sqrt{s}=7$ TeV}, \textit{Phys.Lett.} \textbf{B718} (2013) 815–840, [arXiv:1206.3949].
\bibitem{4} CMS Collaboration, S. Chatrchyan et al., \textit{Search for new physics in events with same-sign dileptons and $b$-tagged jets in $pp$ collisions at $\sqrt{s}=7$ TeV}, \textit{JHEP} \textbf{1208} (2012) 110,
[arXiv:1205.3933].
\bibitem{5} CMS Collaboration, S. Chatrchyan et al., \textit{Search for supersymmetry in hadronic final states with missing transverse energy using the variables $\alpha_T$ and b-quark multiplicity in pp collisions at 8 TeV}, \textit{Eur.Phys.J.} \textbf{C73} (2013) 2568, [arXiv:1303.2985].
\bibitem{6}  CMS Collaboration, S. Chatrchyan et al., \textit{Search for new physics in events with same-sign dileptons and $b$ jets in $pp$ collisions at $\sqrt{s}=8$ TeV}, \textit{JHEP} \textbf{1303} (2013) 037, [arXiv:1212.6194].
\bibitem{7} CMS Collaboration, S. Chatrchyan et al., \textit{Inclusive search for a vector-like T quark with charge 2/3 in pp collisions at $\sqrt{s}=8$ TeV}, \textit{Phys.Lett.} \textbf{B729} (2014) 149–171,
[arXiv:1311.7667].


\end{thebibliography}

\noindent






% \end{thebibliography}

\end{document}